\documentclass[%
preprint,
showpacs,
 amsmath,amssymb,
 aps,
]{revtex4-1}

\usepackage{graphicx}% Include figure files
\usepackage{dcolumn}% Align table columns on decimal point
\usepackage{bm}% bold math
\begin{document}

%\preprint{APS/123-QED}

\title{Dust ion acoustic solitary structures in presence of isothermal positrons}% Force line breaks with \\
%\thanks{asheshpaulju@gmail.com\\animesh.d.d@gmail.com\\abandyopadhyay1965@gmail.com}%

\author{Ashesh Paul}
\author{Anup Bandyopadhyay}
\thanks{abandyopadhyay1965@gmail.com}%
\affiliation{ Department of Mathematics, Jadavpur University, Kolkata
- 700032, India.}%
\author{Animesh Das}
\affiliation{ B. N. S. U. P. School, Howrah - 711 315, India.}%
\date{\today}% It is always \today, today,

\begin{abstract}
\noindent The Sagdeev potential technique has been employed to study the dust ion acoustic solitary waves and double layers in an unmagnetized collisionless dusty plasma consisting of negatively charged static dust grains, adiabatic warm ions, and isothermally distributed electrons and positrons. A computational scheme has been developed to draw the qualitatively different compositional parameter spaces or solution spaces showing the nature of existence of different solitary structures with respect to any parameter of the present plasma system. The qualitatively distinct solution spaces give the overall scenario regarding the existence of different solitary structures. The present system supports both positive and negative potential double layers. The negative potential double layer always restricts the occurrence of negative potential solitary waves, i.e., any sequence of negative potential solitary waves having monotonically increasing amplitude converges to a negative potential double layer.  However, there exists a parameter regime for which the positive potential double layer is unable to restrict the occurrence of positive potential solitary waves. As a result, in this region of the parameter space, there exist solitary waves after the formation of positive potential double layer, i.e., positive potential supersolitons have been observed. But the amplitudes of these supersolitons are bounded. A general theory for the existence of bounded supersolitons has been discussed analytically by imposing the restrictions on the Mach number. For any small value of positron concentration, there is no effect of very hot positrons on the dust ion acoustic solitary structures. The qualitatively different solution spaces are capable of producing new results for the formation of solitary structures.
\end{abstract}

\maketitle

%\tableofcontents

\section{Introduction}
Theoretical and experimental studies of the nonlinear dynamics of Ion Acoustic (IA) waves have received a great deal of attention for several decades. Using a mechanical analogy, \citet{sagdeev66} established that nonlinear IA waves can exist in the form of periodic or solitary waves. On the other hand, in the same year, \citet{washimi66} used reductive perturbation method to investigate the small amplitude ion acoustic solitary waves in a cold plasma. Subsequently, the nonlinear theory of IA waves was developed by a number of authors \cite{schamel72,tagare73,tran74,das75,watanabe78,watanabe84,bharuthram86,tagare86a,tagare86b,verheest88,baboolal88,das89,baboolal89}. Experimental results of \citet{ikezi70}, \citet{nakamura84}, \citet{nakamura85a}, \citet{nakamura85b}, \citet{nishida86}, \citet{nakamura87} and \citet{cooney91} confirmed the existence of IA waves.

There has been considerable interest in studying ion acoustic solitons and double layers in electron-positron-ion (e-p-i) plasmas as such plasmas are found in supernovas, pulsar environments, cluster explosions, active galactic nuclei etc. \citet{popel95} considered the nonlinear propagation of IA waves in cold plasma consisting of cold ions, and isothermally distributed electrons and positrons. They have found the existence of compressive solitons only. In this paper, they reported that the presence of the positron can result in reduction of the IA soliton amplitudes. Subsequently, the properties of the IA solitary structures in different electron-positron-ion (e-p-i) plasmas have been investigated by a number of authors \cite{nejoh96,nejoh97,salahuddin02,alinejad06,esfandyari,chawla10,sahu12,el-tantawy12,mishra07,mahmood08,boubakour09,dubinov09,sabry09a,sabry09b,tribeche09,chatterjee10a,sahu11,baluku11,ghosh11,tribeche11,shah12,jain13,hussain13,kaladze14,saha14}.

However, in astrophysical environments highly (negative/positive) charged micronsize impurities or dust particulates are observed in addition to earlier mentioned e-p-i plasma. The presence of dust grains having large masses introduces several new aspects in the properties of the nonlinear waves and coherent structures \cite{rao90,shukla92,verheest92,barkan96,mamun96a,mamun96b,pieper96,shukla01,shukla09a,shukla99,shukla03}. Depending on different time scales, there can exist two or more acoustic waves in a typical dusty plasma. Dust Acoustic  (DA)  and  Dust Ion Acoustic  (DIA)  waves are  two  such  acoustic  waves. \citet{shukla92} were the first to show that due to the quasi-neutrality condition $ n_{e0} + n_{d0}Z_{d}  = n_{i0} $ and the strong inequality $ n_{e0} \ll n_{i0} $ (where $ n_{e0}, n_{i0},$ and $ n_{d0} $ are, respectively, the number density of electrons, ions, and dust particles, and $ Z_{d} $ is the number of electrons residing  on  the  dust  grain  surface),  a  dusty  plasma (with negatively charged static dust grains) supports low-frequency Dust Ion Acoustic (DIA) waves with phase velocity much smaller (larger) than electron (ion) thermal velocity. In the case of  a  long  wavelength  limit,  the  dispersion  relation  of DIA wave is similar to that of IA wave for a plasma with $ n_{e0} = n_{i0} $ and $ T_{i} \ll T_{e} $, where $ T_{i}(T_{e}) $ is the average ion (electron) temperature. Due to the usual dusty plasma approximations ($ n_{e0} \ll n_{i0} $ and $ T_{i}  \simeq T_{e} $), a dusty plasma cannot support the usual IA waves, but a dusty plasma can support the DIA waves of \citet{shukla92}. Thus, DIA waves are basically IA waves modified  by  the  presence  of  heavy  dust  particulates. The  theoretical  prediction  of  \citet{shukla92} was supported by a number of laboratory experiments \cite{barkan96,merlino98,nakamura01}. The nonlinear theory of DIA waves in different dusty plasma systems has been investigated by \citet{bharuthram92} , \citet{nakamura99} , \citet{luo99} , \citet{mamun02a} , \citet{shukla03} ,  \citet{verheest05} , \citet{sayed08} , \citet{alinejad10b}, \citet{baluku10} , \citet{das12}.

It is well known that a fully ionized universe contains electrons, positrons, ions and micron-sized charged dust grains \cite{evans94,spitzer78}. The presence of e-p-i-d plasma has been detected in active galactic nuclei, pulsar magnetospheres, inter-stellar clouds, supernova environments as well as in laboratory  experiments  of  cluster  explosions  by  intense  laser beams \cite{shukla04,tajima97,weinberg72,higdon09,Shukla97,cho00}. Therefore the nonlinear dynamics of the propagation of waves in this plasma system has emerged as an interesting field to explore.  For the first time, \citet{ghosh08} investigated the nonlinear propagation of small but finite amplitude ion acoustic solitons and double layers in a collisionless unmagnetized e-p-i-d  plasma consisting of cold ions, negatively charged static dust particulates and Boltzmann distributed electrons and positrons. They have used the reductive perturbation method to derive the  Korteweg-de Vries (KdV) equation and the modified KdV equations for the ion acoustic waves in the e-p-i-d plasma system. Using KdV equation, they proved that this e-p-i-d plasma system supports both compressive and rarefactive solitons if the parameters of the system satisfy certain conditions. Using the modified KdV equation, they have derived the necessary conditions for the existence of weak double layers. Employing the reductive perturbation technique, \citet{el-tantawy11a} investigated the ion acoustic solitary structures in a collisionless unmagnetized four-component e-p-i-d plasma consisting of warm ions, superthermal electrons and positrons, and negatively charged static dust impurities. Using Bernoulli's pseudo-potential method, \citet{dubinov12} elaborated the nonlinear theory of dust ion acoustic waves  in a collisionless unmagnetized four-component e-p-i-d plasma consisting of warm ions, isothermal electrons and positrons, and negatively charged static dust impurities. They have found the existence of four different types of physically realizable nonlinear ion acoustic waves: a subsonic periodic wave, a large amplitude subsonic periodic wave (superlinear wave), supersonic rarefactive and compressive solitary waves.

In the above mentioned papers of nonlinear DIA or IA waves in e-p-i-d plasmas, the authors have not considered the following interesting points. From the investigations of DIA waves in a collisionless unmagnetized three-component e-i-d plasma consisting of warm ions, isothermal or nonthermal electrons and negatively charged static dust impurities, we have seen from the article of \citet{das12} that the system supports the Negative Potential Solitary Waves (NPSWs) \textbf{for all} $M>M_{c}$, where $M_{c}$ is the lower bound of the Mach number $M$ for the existence of DIA solitary structures, i.e., DIA solitary structures start to exist \textbf{for all} $M>M_{c}$. From this simple conclusion of \citet{das12}, we have the following questions regarding the existence of different DIA solitary structures when an amount of positrons is injected in the system: Can we restrict the occurrence of NPSWs after injecting positron in the system? Does the system support Negative Potential Double Layer (NPDL)? Does the NPSW disappear from the system if the concentration of positron increases? Does the system support Positive Potential Double Layer (PPDL)?  Does the system support the coexistence of double layers of both polarities? Does the system support supersolitons? (i.e., Is there any soliton after the formation of double layer? ) Is the amplitude of the supersoliton bounded?  What is the condition on the Mach number for the existence of bounded supersolitons (supersolitons of bounded amplitude)? or, Can we impose any restriction on the Mach number to get bounded supersolitons? Is there any connection between the occurrence of suppersolitons and the occurrence of NPSWs and / or Positive Potential Solitary Waves (PPSWs)? Does there exist a critical value of positron concentration for which the system supports only positive potential solitary structures (PPSWs and PPDLs)? Does there exist a critical value of positron concentration for which the system supports only Positive Potential Solitary Waves (PPSWs)? What happens if very hot positrons are injected in the system? or Is there any qualitative difference in the DIA solitary structures if very hot positrons are injected in the system? To answer all these questions, we reconsider the same problem of \citet{das12} by making the following changes: instead of taking three component e-i-d plasma, a collisionless unmagnetized four component  e-p-i-d plasma has been considered in which electrons and positrons both are isothermally distributed, i.e., in the present investigation, we reconsider the problem of \citet{ghosh08} in the following directions: (i) the equation of pressure for ion fluid is taken into account to include the effect of ion temperature, (ii) instead of considering the reductive perturbation method, \citet{sagdeev66} pseudo-potential technique has been employed to investigate arbitrary amplitude DIA solitary waves, double layers and supersolitons in unmagnetized collisionless e-p-i-d plasma, (iii) the investigations have been made with the help of the qualitatively different compositional parameter spaces instead of considering particular values of the parameters involved in the system, (iv) a general theory has been given on the existence of bounded supersolitons which describes the condition on the Mach number for the occurrence of bounded supersolitons, (v) a thorough investigation on existence of DIA solitary structures has been presented through the compositional parameter spaces, giving the special emphasis on the existence of bounded supersolitons.

Four basic parameters of the present e-p-i-d plasma system are $p$,  $\mu$,  $\sigma_{pe}$ and $\sigma_{ie}$ which are, respectively, the ratio of unperturbed number density of  positrons to that of the total unperturbed number density of positive charge, the ratio of unperturbed number density of electrons to that of the total unperturbed number density of negative charge, the ratio of the average temperature of positrons to that of electrons and the ratio of average temperature of ions to that of electrons. The studies of the DIA solitary structures of the present e-p-i-d plasma system have been made over the entire physically admissible values of $p$,  $\mu$,  $\sigma_{pe}$ and $\sigma_{ie}$.

The present paper is organized as follows: In \S \ref{sec:basic_eqn}, the basic equations are given. The derivation and the mechanical analogy of the energy integral are given in \S \ref{sec:energy_int}. Conditions for the existence of solitary wave and double layer solutions are also given in this section.  A general theory for the formation of bounded supersolitons has been presented in \S \ref{sec:L_U_Bounds} by imposing the restrictions on the Mach number. The lower bound and the upper bound of the Mach number for the existence of different solitary structures have been determined in this section. With the help of the analytical theory discussed in \S \ref{sec:L_U_Bounds}, a computational scheme has been developed in \S \ref{sec:solution_spaces} to draw the qualitatively different compositional parameter spaces with respect to any parameter of the system. In \S \ref{sec:solution_spaces}, DIA solitary structures of the present plasma system have been thoroughly presented with the help of the qualitatively distinct compositional parameter spaces, giving special emphasis on the bounded positive potential supersolitons.  Finally, a brief summary and discussions have been given in \S \ref{sec:Summary_Discussions}.

\section{\label{sec:basic_eqn}Basic Equations}
The following are the governing equations describing the non-linear behaviour of dust ion acoustic waves  propagating  along  x-axis  in  collisionless  unmagnetized  dusty  plasma  consisting of adiabatic warm ions,  negatively  charged  immobile  dust  grains, and isothermally distributed electrons and positrons:
\begin{eqnarray}\label{continuity}
\frac{\partial n_{i}}{\partial t}+\frac{\partial}{\partial x}(n_{i}u_{i})=0,
\end{eqnarray}
\begin{eqnarray}\label{momentum}
n_{i}m_{i}\bigg(\frac{\partial u_{i}}{\partial t}+u_{i}\frac{\partial u_{i}}{\partial x}\bigg)+\frac{\partial p_{i}}{\partial x}+n_{i}q_{i}\frac{\partial \phi}{\partial x}=0,
\end{eqnarray}
\begin{eqnarray}\label{pressure}
\frac{\partial p_{i}}{\partial t}+u_{i}\frac{\partial p_{i}}{\partial x}+\gamma p_{i} \frac{\partial u_{i}}{\partial x}=0,
\end{eqnarray}
\begin{eqnarray}\label{poisson}
\frac{\partial^{2} \phi}{\partial x^{2}}=-4\pi e(n_{i}-n_{e}+n_{p}-Z_{d}n_{d}).
\end{eqnarray}

\noindent Here $ n_{i}$, $n_{e}$, $n_{p}$, $n_{d}$, $u_{i}$, $p_{i}$, $\phi$, $x $ and $ t $ are, respectively, ion number density, electron number density, positron number density, dust particle number density, ion fluid velocity, ion fluid pressure, electrostatic potential, spatial variable and time, $ \gamma(=3) $ is the adiabatic index, $ m_{i} $ is the mass of ion fluid, $ Z_{d} $ is the number of negative unit charges residing on the dust grain surface and $ e $ is the charge of an electron.

The above four equations are supplemented by
\begin{eqnarray}\label{ne}
\frac{n_{e}}{n_{e0}}=exp\bigg[\frac{\phi}{\Phi}\bigg]~,~
\frac{n_{p}}{n_{p0}}=exp\bigg[-\frac{\phi}{\sigma_{pe} \Phi}\bigg],
\end{eqnarray}
and the charge neutrality condition
\begin{eqnarray}\label{charge_neutrality}
n_{i0}+n_{p0}=n_{e0}+Z_{d}n_{d0},
\end{eqnarray}
where $n_{e0}$, $n_{i0}$, $n_{p0}$ and $n_{d0}$ are, respectively, the unperturbed number densities of electron, ion, positron and dust particulate, $\Phi$ and $\sigma_{pe}$ are given by
\begin{eqnarray}\label{Phi}
\Phi=\frac{K_{B}T_{e}}{e}, \sigma_{pe}=\frac{T_{p}}{T_{e}}.
\end{eqnarray}
Here $K_{B}$ is the Boltzmann constant, $T_{e}$ and $T_{p}$ are the average temperatures of electrons and positrons respectively.
%%%%%%%%%%%%%%%%%%%%%%%%%%%%%%%%%%%%%%%%%%%%%%%%%%%%%%%%%%%%%%%%%%%%%%%%%%%%%%%%%%%%%%%%%%%%%%%%%%%%%%%%%%%%%%%%%%%%%%%%%%%%%%%%%%%%%%%%%%%%%%
%%%%%%%%%%%%%%%%%%%%%%%%%%%%%%%%%%%%%%%%%%%%%%%%%%%%%%%%%%%%%%%%%%%%%%%%%%%%%%%%%%%%%%%%%%%%%%%%%%%%%%%%%%%%%%%%%%%%%%%%%%%%%%%%%%%%%%%%%%%%%%
\section{\label{sec:energy_int} Energy Integral}
The linear dispersion relation of the DIA wave for the present dusty plasma system can be written as
\begin{eqnarray}\label{dispersion_relation}
\frac{\omega}{k}=c_{D}\sqrt{\frac{1+\frac{\gamma\sigma_{ie}}{M_{s}^{2}}k^{2}\lambda_{D}^{2}}{1+k^{2}\lambda_{D}^{2}}},c_{D}=C_{s}M_{s}
\end{eqnarray}
where $\omega$ and $k$ are respectively the wave frequency and wave number of the plane wave perturbation, and
\begin{eqnarray}\label{cD}
C_{s}=\sqrt{\frac{K_{B}T_{e}}{m_{i}}}, M_{s}=\sqrt{\gamma\sigma_{ie}+\frac{(1-p)\sigma_{pe}}{p+\mu \sigma_{pe}}},
\end{eqnarray}
\begin{eqnarray}\label{lambda_D_square}
\frac{1}{\lambda_{D}^{2}}=\frac{1}{\lambda_{Dp}^{2}}+\frac{1}{\lambda_{De}^{2}}~,
\end{eqnarray}
\begin{eqnarray}\label{lambda_Dp}
\lambda_{Dp}^{2}=\frac{K_{B}T_{p}}{4\pi e^{2}n_{p0}}~,~
\lambda_{De}^{2}=\frac{K_{B}T_{e}}{4\pi e^{2}n_{e0}},
\end{eqnarray}
\begin{eqnarray}\label{sigma_ie}
\sigma_{ie}=\frac{T_{i}}{T_{e}}~,~\sigma_{pe}=\frac{T_{p}}{T_{e}}~,~
p=\frac{n_{p0}}{n_{0}}~,~\mu=\frac{n_{e0}}{n_{0}},
\end{eqnarray}
\begin{eqnarray}\label{n0}
n_{0}=n_{i0}+n_{p0}=n_{e0}+Z_{d}n_{d0}.
\end{eqnarray}

Now for long-wave length plane wave perturbation, i.e., for $ k \rightarrow 0 $, from linear dispersion
relation (\ref{dispersion_relation}), we have,
\begin{eqnarray}
\lim_{k \to 0}\frac{\omega}{k} =c_{D} \mbox{   and   } \lim_{k \to 0}\frac{d\omega}{dk} = c_{D}
\end{eqnarray}
and consequently the dispersion relation (\ref{dispersion_relation}) shows that the linearized velocity of the DIA wave in the present plasma system is $ c_{D} $ with $ \lambda_{D} $ as the Debye length.

To  study  the  arbitrary  amplitude  time  independent DIA solitary waves and double layers, we make all the dependent variables depend only on a single variable $ \xi=x-Ut $ where $ U $ is independent of $ x $ and $ t $. Thus, in the wave frame moving with a constant velocity $U$ the equations (\ref{continuity})-(\ref{poisson}) can be put in the following form
\begin{eqnarray}\label{modified_continuity}
\frac{d}{d\xi}\bigg\{(U-u_{i})\frac{n_{i}}{n_{i0}}\bigg\}=0,
\end{eqnarray}
\begin{eqnarray}\label{modified_momentum}
\frac{d}{d\xi}\big(-Uu_{i}+\frac{u_{i}^{2}}{2}+C_{s}^{2}\frac{\phi}{\Phi}\big)
 +\sigma_{ie}C_{s}^{2}\frac{n_{i0}}{n_{i}}\frac{d}{d\xi}\big(\frac{p_{i}}{P}\big)=0,
\end{eqnarray}
\begin{eqnarray}\label{modified_pressure}
\frac{d}{d\xi}\big\{(U-u_{i})\big(\frac{p_{i}}{P}\big)^{1/\gamma}\big\}=0,
\end{eqnarray}
\begin{eqnarray}\label{modified_poisson}
\frac{d^{2}}{d\xi^{2}}\bigg(\frac{\phi}{\Phi}\bigg) = \frac{M_{s}^{2}-\gamma \sigma_{ie}}{\lambda_{D}^{2}}\Bigg[\frac{1-\mu}{1-p}+ \frac{\mu}{1-p}\frac{n_{e}}{n_{e0}}-\frac{p}{1-p}\frac{n_{p}}{n_{p0}}-\frac{n_{i}}{n_{i0}}\Bigg].
\end{eqnarray}
Here $ P=n_{i0}K_{B}T_{i} $, $ n_{i0} $ is the unperturbed ion number density and $ T_{i} $ is the average temperature of ions.

Using the boundary conditions,
\begin{eqnarray}\label{boundary_condition}
\big(\frac{n_{i}}{n_{i0}},\frac{p_{i}}{P},u_{i},\phi,\frac{d\phi}{d\xi}\big)\rightarrow \big(1,1,0,0,0\big)\mbox{    as    }  |\xi|\rightarrow \infty
\end{eqnarray}
 and solving (\ref{modified_continuity}), (\ref{modified_momentum}), and (\ref{modified_pressure}), we get a quadratic equation  for  $ n_{i}^{2} $, and  the  solution of the final equation of $ n_{i} $ can be put in the following form:
\begin{eqnarray}\label{ni_square}
\frac{n_{i}}{n_{i0}}=N_{i}=\frac{(U/C_{s})\sqrt{2}}{\sqrt{\frac{\Phi_{U}}{\Phi}-\frac{\phi}{\Phi}}+\sqrt{\frac{\Psi_{U}}{\Phi}-\frac{\phi}{\Phi}}},
\end{eqnarray}
where
\begin{eqnarray}\label{phi_u}
\frac{\Phi_{U}}{\Phi} = \frac{1}{2}\big(\frac{U}{C_{s}}+\sqrt{3\sigma_{ie}}\big)^{2},
\frac{\Psi_{U}}{\Phi} = \frac{1}{2}\big(\frac{U}{C_{s}}-\sqrt{3\sigma_{ie}}\big)^{2}
\end{eqnarray}
Now  integrating  (\ref{modified_poisson})  with  respect  to  $ \phi $  and  using the  boundary  conditions  (\ref{boundary_condition}),  we  get  the  following equation  known  as  energy  integral  with  $ W(\phi) $  as  the Sagdeev potential or pseudo-potential:
\begin{eqnarray}\label{energy_integral}
\frac{1}{2}\bigg(\frac{d\phi}{d\xi}\bigg)^{2}+W(\phi)=0
\end{eqnarray}
where
\begin{eqnarray}\label{V_phi}
W(\phi) = \frac{\Phi^{2}}{\lambda_{D}^{2}}(M_{s}^{2}-3 \sigma_{ie})\Bigg[~~W_{i}+\frac{p \sigma_{pe}}{1-p}W_{p}  - \frac{\mu}{1-p} W_{e}-\frac{1-\mu}{1-p}W_{d}\Bigg],
\end{eqnarray}
\begin{eqnarray}\label{V_i}
W_{i} = \frac{U^{2}}{C_{s}^{2}}+\sigma_{ie}- N_{i}\big(\frac{U^{2}}{C_{s}^{2}}+3\sigma_{ie}
-2\frac{\phi}{\Phi}-2\sigma_{ie}N_{i}^{2}\big),
\end{eqnarray}
\begin{eqnarray}\label{V_e}
W_{e} = exp\bigg(\frac{\phi}{\Phi}\bigg)-1,
W_{p} = 1-exp\bigg(-\frac{\phi}{\sigma_{pe} \Phi}\bigg),
\end{eqnarray}
\begin{eqnarray}
W_{d} &=& \frac{\phi}{\Phi}.
\end{eqnarray}
The energy integral (\ref{energy_integral}) can be regarded as the equation of energy of a particle of unit mass moving along a straight line whose position is $\phi$ at time $\xi$ with velocity $\frac{d \phi}{d \xi}$. The first term of the energy integral (\ref{energy_integral}) can be regarded as the kinetic energy of the particle whereas the second term of the energy integral (\ref{energy_integral}) can be regarded as the potential energy of the same particle at that instant. As the kinetic energy is a non-negative quantity, $W(\phi) \leq 0$ for the entire motion of the particle. Differentiating the energy integral (\ref{energy_integral}) with respect to $\phi$, we get the following equation
\begin{eqnarray}\label{energy_motion}
\frac{d^{2} \phi}{d \xi^{2}}+W'(\phi)=0, W'(\phi)=\frac{dW}{d \phi}
\end{eqnarray}
The above equation shows that the particle of unit mass is under the action of the force $-W'(\phi)$ and this force is attracting one, i.e., directed towards the point $\phi=0$, i.e., $\phi=0$ is the centre of the force if $W'(\phi)>0$ for $\phi>0$ or $W'(\phi)<0$ for $\phi<0$. Now, if $W(0)=W'(0)=0$, then both the velocity and the force acting on the particle at $\phi =0$ are simultaneously equal to zero and consequently, the particle is in equilibrium at $\phi=0$. Let $\phi=0$ be an equilibrium position of the particle associated with the energy integral (\ref{energy_integral}). The nature of the equilibrium point $\phi=0$ depends on the sign of $W''(0)$. In fact, we have the following cases depending on the sign of $W''(0)$ : $W''(0)>0$ and $W''(0)<0$. The case $W''(0)>0$ implies that $\phi=0$ is a position of stable equilibrium but this case is not possible because for this case the radius of curvature of the curve $W(\phi)$ at $\phi=0$ is positive and consequently the curve $W(\phi)$ is concave with respect to the $\phi$ - axis in a neighbourhood of $\phi=0$ which again implies that $W(\phi)>0$ for any $\phi$ (except the point $\phi=0$) lying within a very small neighbourhood of $\phi=0$. But $W(\phi)>0$ is not possible for the particle which has been associated with the energy integral (\ref{energy_integral}) as we have seen earlier that $W(\phi)\leq 0$ for the entire motion of the particle. Next we consider the case $W''(0)<0$. For this case, $\phi=0$ is a position of unstable equilibrium of the particle and $W(\phi)$ is convex with respect to the $\phi$ - axis in a neighbourhood of $\phi=0$ which again implies that $W(\phi)<0$ for any $\phi$ (except the point $\phi=0$) lying within a very small neighbourhood of $\phi=0$. So, if the particle placed at $\phi=0$ be slightly displaced towards the positive (negative) direction of $\phi$ - axis, it moves away from its position of unstable equilibrium and it continues its motion until its velocity is equal to zero, i.e., until $\phi$ takes the value $\phi_{m}$ such that $W(\phi_{m})=0$ for $\phi_{m}>0$ ($\phi_{m}<0$). Now, if $W'(\phi_{m})>0$ for $\phi_{m}>0$ ($W'(\phi_{m})<0$ for $\phi_{m}<0$), then although the velocity of the particle at $\phi=\phi_{m}$ vanishes, the force acting on the particle at $\phi=\phi_{m}$ is directed towards the point $\phi=0$ and consequently, the particle will come back again at $\phi=0$. Therefore, if $W(0)=W'(0)=0$, $W''(0)<0$, $W(\phi_{m})=0$ and $W'(\phi_{m})>0$ for $\phi_{m}>0$ ($W'(\phi_{m})<0$ for $\phi_{m}<0$) then the energy integral (\ref{energy_integral}) can be regarded as the equation of energy of the oscillatory motion of a particle of unit mass, i.e., we have an oscillation of the particle within $0<\phi<\phi_{m}$ ($\phi_{m} < \phi <0$). As this oscillation takes place in a wave frame moving with a velocity $U$, this oscillation propagates with velocity $U$ and in this wave frame this oscillation is known as positive (negative) potential soliton. From the mechanical analogy of the energy integral (\ref{energy_integral}) as mentioned above, \citet{sagdeev66} established that for the existence of a Positive Potential Solitary Wave (Negative Potential Solitary Wave) solution of the energy integral (\ref{energy_integral}), the following three conditions must be simultaneously satisfied.
\begin{description}
  \item[Sa :] $\phi=0$ is the position of unstable equilibrium of the particle, i.e., $W(0)=W'(0)=0$ and $W''(0)<0$.
  \item[Sb :] $W(\phi_{m}) = 0$, $W'(\phi_{m}) > 0$ for some $\phi_{m} > 0$ ($W'(\phi_{m}) < 0$ for some $\phi_{m} < 0$). This condition is responsible for the oscillation of the particle within the interval $\min\{0,\phi_{m}\}<\phi<\max\{0,\phi_{m}\}$.
  \item[Sc :] $W(\phi) < 0$ for all $0 <\phi < \phi_{m}$ ($W(\phi) < 0$ for all $\phi_{m} < \phi < 0$). This condition is necessary to define the energy integral (\ref{energy_integral}) within the interval $\min\{0,\phi_{m}\}<\phi<\max\{0,\phi_{m}\}$.
\end{description}

On the other hand, if $W'(\phi_{m}) = 0$ along with the condition $W(\phi_{m}) = 0$ then the velocity and the force acting on the particle at $\phi =\phi_{m}$ are simultaneously equal to zero and consequently the particle will not be reflected back again at $\phi=0$. In this case instead of soliton solution Energy Integral (\ref{energy_integral}) gives shock-like solution which is known as double layer solution. If $\phi_{m} > 0$ the double layer is known as positive potential double layer (PPDL) whereas if $\phi_{m} < 0$ the double layer is known as negative potential double layer (NPDL). Therefore, for the existence of a Positive Potential Double Layer (Negative Potential Double Layer) solution of the energy integral (\ref{energy_integral}), the following three conditions must be simultaneously satisfied.
\begin{description}
  \item[Da :] $\phi=0$ is the position of unstable equilibrium of the particle, i.e., $W(0)=W'(0)=0$ and $W''(0)<0$.
  \item[Db :] $W(\phi_{m}) = 0$, $W'(\phi_{m}) = 0$, $W''(\phi_{m}) < 0$ for some $\phi_{m} > 0$ ($\phi_{m} < 0)$. This condition actually states that the particle cannot be reflected back again at $\phi = 0$.
  \item[Dc :] $W(\phi) < 0$ for all $0 <\phi < \phi_{m}$ ($W(\phi) < 0$ for all $\phi_{m} < \phi < 0$). This condition is necessary to define the energy integral (\ref{energy_integral}) within the interval $\min\{0,\phi_{m}\}<\phi<\max\{0,\phi_{m}\}$.
\end{description}

Here, we have not discussed the case for the existence of solitary structures (solitons and double layers) when $W''(0)=0$. Following \citet{das12mc} one can discuss this case for this particular problem.

Therefore, the  necessary  conditions  for  the  existence  of  solitary structures  of  the  energy  integral  (\ref{energy_integral})  are  $W (0) = 0$, $W '(0)  =  0$,  and $ W ''(0)  <  0$.  It  can  be  easily  checked that $W (0) = 0$, $W '(0)  =  0$, and the condition $ W ''(0)  <  0$ gives $ U  > c_{D} $,  and  consequently  the  greatest lower bound (glb) of the Mach number ($ M = U/c_{D}$, which has been explicitly discussed by \citet{dubinov09a}) for  the  existence  of  solitary  structures  of  the  energy integral (\ref{energy_integral}) is $ M_{c}  = 1 $, i.e., the solitary structures start to exist for $ U > c_{D}   (\Leftrightarrow  M  > M_{c}  = 1)$.  So  we  have  introduced  the following dimensionless quantities: $ \bar{x}=x/\lambda_{D} $, $ \overline{\xi}=\xi/\lambda_{D} $, $ \overline{t}=t/(\lambda_{D}/c_{D}) $, $ \overline{u_{i}}=u_{i}(\lambda_{D}/c_{D})/\lambda_{D} = u_{i}/c_{D} $, $ \overline{\phi}=\phi/\Phi $, $ \overline{p_{i}}=p_{i}/P $, $ \overline{n_{s}}=n_{s}/n_{0} $. Now we note the following fact: $\xi=x-Ut=\lambda_{D}\overline{x}-U\frac{\lambda_{D}}{c_{D}}\overline{t}=\lambda_{D}[\overline{x}-\frac{U}{c_{D}}\overline{t}]
=\lambda_{D}[\overline{x}-M\overline{t}] \Leftrightarrow \frac{\xi}{\lambda_{D}}=\overline{x}-M\overline{t} \Leftrightarrow \overline{\xi}=\overline{x}-M\overline{t}$, i.e., here spatial coordinate is normalized by $\lambda_{D}$ and the time is normalized by $\frac{\lambda_{D}}{c_{D}}$. Then with respect to these dimensionless quantities, the energy integral can be simplified as follows, where we drop 'overline' on both independent and dependent variables:

\begin{eqnarray}\label{energy_integral_1}
\frac{1}{2}\bigg(\frac{d\phi}{d\xi}\bigg)^{2}+V(\phi)=0,
\end{eqnarray}
where
\begin{eqnarray}\label{V_phi_1}
V(\phi) = (M_{s}^{2}-3\sigma_{ie}) \Bigg[~V_{i} +\frac{p }{1-p} \sigma_{pe} V_{p}-\frac{\mu}{1-p}V_{e}-\frac{1-\mu}{1-p}V_{d}\Bigg],
\end{eqnarray}
\begin{eqnarray}\label{V_i_1}
V_{i} = M^{2}M_{s}^{2}+\sigma_{ie}-N_{i}(M^{2}M_{s}^{2}+3\sigma_{ie}-2\phi-2\sigma_{ie}N_{i}^{2}),\nonumber
\end{eqnarray}
\begin{eqnarray}\label{N_i_1}
N_{i}=\frac{n_{i}}{n_{i0}}=\frac{MM_{s}\sqrt{2}}{(\sqrt{\Phi_{M}-\phi}+\sqrt{\Psi_{M}-\phi})}
\end{eqnarray}
\begin{eqnarray}\label{Phi_M_1}
\Phi_{M} &=& \frac{1}{2}\bigg(MM_{s}+\sqrt{3\sigma_{ie}}\bigg)^{2}, \\
 \Psi_{M} &=& \frac{1}{2}\bigg(MM_{s}-\sqrt{3\sigma_{ie}}\bigg)^{2},
\end{eqnarray}
\begin{eqnarray}\label{V_e_1}
V_{e}=e^{\phi}-1,V_{p} = 1-e^{-\phi/\sigma_{pe}},
V_{d}=\phi.
\end{eqnarray}
The equations (\ref{energy_integral}) and (\ref{energy_integral_1}) are dynamically equivalent and consequently, the qualitative behaviour of solitary structures determined by these two equations are same. In fact, one can take any equation equivalent to the energy integral (\ref{energy_integral}) to discuss the qualitative behaviour of the solitary structures, but in order to avoid quantitative inaccuracies, the general prescription is to normalize length by $\lambda_{D}$ and time by $\lambda_{D}/c_{D}$, where $\lambda_{D}$ is the actual Debye length and $c_{D}$ is the actual acoustic speed of the plasma system [\citet{dubinov09a}]. One can obtain $c_{D}$ by considering the linear dispersion relation as prescribed by \citet{dubinov09a}. So, we can take the energy integral (\ref{energy_integral_1}) to investigate the solitary structures of the present plasma system.

From the expression of $ N_{i} $ as given by (\ref{N_i_1}), we see that $ N_{i} $ is  well - defined  if and only if both $ \Phi_{M}-\phi $  and $ \Psi_{M}-\phi $ are real and non-negative. Now the conditions $ \Phi_{M}-\phi \geq 0$  and $ \Psi_{M}-\phi \geq 0$ hold simultaneously if and only if $ \phi \leq Min\{\Phi_{M}, \Psi_{M} \} $. As $ \Psi_{M} \leq \Phi_{M}$, the equation  (\ref{N_i_1}) gives theoretically valid expression of $ N_{i} $ if and only if $ \phi \leq \Psi_{M} $.  Therefore, $\phi$ is restricted by the inequality: $ \phi \leq \Psi_{M} = (MM_{s}-\sqrt{3\sigma_{ie}})^{2}/2 $. In fact, we have used this inequality, viz., $ \phi \leq \Psi_{M} $ to find the upper bound of the Mach number for the existence of PPSWs.

\section{\label{sec:L_U_Bounds} Analytical theory for the formation of supersolitons}

In this section, our main aim is to find the Mach number for the existence of supersolitons. Before going to investigate the existence of supersolitons analytically, first of all, we note the following facts:

For  the  existence  of  solitary structures  of  the  energy  integral  (\ref{energy_integral_1}), we have $V (0) = 0$, $V '(0)  =  0$,  and $ V ''(0)  <  0$.  It  can  be  easily  checked that $V (0) = 0$, $V '(0)  =  0$, and the condition $ V ''(0)  <  0$ gives $M>M_{c}=1$. Therefore, the solitary structures start to exist for $M > M_{c}=1$, i.e., $M=M_{c}=1$ is the lower bound of the Mach number $M$ for the  existence  of  solitary structures.

If there exist solitons after the formation of double layer of same polarity for increasing values of the Mach number then the solitons are known as supersolitons. In other words, suppose we have a double layer solution at $M=M_{DL}$ of some polarity and for definiteness, let us assume that we have a positive potential double layer solution at $M=M_{DL}$. If there exists a Mach number $M_{P}(>M_{DL})$ such that we have a PPSW at $M=M_{P}(>M_{DL})$, then the PPSW at $M=M_{P}(>M_{DL})$ is called a positive potential supersoliton. Similarly, one can define negative potential supersolitons. Existence of supersolitons for some particular values of the parameters involved in different plasma system have been recently reported by many authors \cite{verheest09,baluku10a,verheest11,das12,dubinov12a,dubinov12b,dubinov12c,maharaj13,hellberg13,verheest13a,verheest13b,verheest13c,verheest14,verheest14a,rufai14,lakhina14,singh15}. In the above investigations, except \citet{das12}, the authors have studied the existence of supersolitons for some particular values of the parameters involved in different plasma systems but \citet{das12} have studied the unbounded negative potential supersolitons through the compositional parameter space. They have clearly pointed out the existence region of the unbounded negative potential supersolitons but instead of `supersolitons' they have used the term `dias type solitons'.

From the physical interpretation of the energy integral for the existence of the double layer of any polarity, we have found that the existence of the double layer of any polarity always implies that there must exist a sequence of solitary waves of same polarity having monotonically increasing amplitude converging to the double layer solution, i.e., the amplitude of the double layer solution acts as an exact upper bound or least upper bound ($lub$) of the amplitudes of at least one sequence of solitary waves of same polarity. Therefore, if the double layer solution exists then this double layer solution is a limiting structure of a sequence of solitary waves of same polarity. On the other hand, any sequence of solitary waves ends with a double layer of same polarity if it exists. So, double layer solution plays an important role to restrict the occurrence of at least one sequence of solitary waves of same polarity. More specifically, if $M=M_{PPDL}$ ($M=M_{NPDL}$) corresponds to the positive (negative) potential double layer then there must exist positive (negative) potential solitary waves for any $M$ restricted by the inequality $M_{c}<M<M_{PPDL}$ ($M_{c}<M<M_{NPDL}$) and the amplitude of the solitary wave increases with increasing $M$ and these solitary waves end with a double layer solution of same polarity when $M$ assumes the value $M=M_{PPDL}$ for positive potential double layer ($M=M_{NPDL}$ for negative potential double layer). Therefore, double layer solution plays an important role to restrict the occurrence of solitary waves of same polarity. Of course, our aim is to find a Mach number $M_{P}>M_{PPDL}$ ($M_{N}>M_{NPDL}$) such that there exists a PPSW (NPSW) at $M=M_{P}$ ($M=M_{N}$). Now, if the plasma system does not support any PPDL (NPDL) for any given set of values of the parameters of the system, then there is no question of existence of positive (negative) potential supersolitons. So, it is important to investigate whether the system supports any double layer solution. Following \citet{das12}, we shall analytically investigate the existence of double layer solution of the energy integral (\ref{energy_integral_1}).

For the existence of a double layer solution of the energy integral (\ref{energy_integral_1}), we must have a non-zero
$ \phi $ $ (\phi \neq 0) $ such that the following conditions are simultaneously satisfied:
\begin{eqnarray}\label{inequality_5}
V(\phi) = 0, V'(\phi) = 0, V''(\phi) < 0
\end{eqnarray}
Using equations (\ref{V_phi_1}) –- (\ref{V_e_1}), the first equation, the second equation, and the third inequality of (\ref{inequality_5}) can be written, respectively, as
\begin{eqnarray}\label{V_phi_2}
V(\phi) \equiv (1-N_{i})M^{2}M_{s}^{2} +\sigma_{ie}  -N_{i}(3\sigma_{ie}-2\phi-2\sigma_{ie}N_{i}^{2})-S = 0,
\end{eqnarray}
\begin{eqnarray}\label{V_dash_phi_}
V'(\phi) \equiv N_{i}-\frac{dS}{d\phi} = 0,
\end{eqnarray}
\begin{eqnarray}\label{V_doubledash_phi_}
V''(\phi) \equiv  \frac{d}{d\phi}\bigg(N_{i}-\frac{dS}{d\phi}\bigg) < 0,
\end{eqnarray}
where
\begin{eqnarray}\label{S}
S = \frac{\mu}{(1-p)}V_{e}+\frac{1-\mu}{(1-p)}V_{d}-\frac{p\sigma_{pe}}{(1-p)}V_{p},
\end{eqnarray}
Eliminating $N_{i}$  from (\ref{V_phi_2}) and (\ref{V_dash_phi_}), we get
\begin{eqnarray}\label{equation_1}
M^{2} &=& \frac{S-\sigma_{ie} +\frac{dS}{d\phi}\bigg[3\sigma_{ie} -2\phi-2\sigma_{ie}\bigg(\frac{dS}{d\phi}\bigg)^{2}\bigg]}{M_{s}^{2}\bigg(1-\frac{dS}{d\phi}\bigg)}.
\end{eqnarray}
It can be easily checked that $ \phi=0 $ if and only if $ dS/d\phi=1 $, and consequently for non-zero $ \phi $, (\ref{equation_1}) can be written as
\begin{eqnarray}\label{equation_2}
M^{2}M_{s}^{2} = h(\phi),
\end{eqnarray}
where
\begin{eqnarray}\label{h_phi}
h(\phi) = \frac{S-\sigma_{ie} +\frac{dS}{d\phi}\big[3\sigma_{ie}-2\phi
-2\sigma_{ie}\big(\frac{dS}{d\phi}\big)^{2}\big]}{\big(1-\frac{dS}{d\phi}\big)}.
\end{eqnarray}
Using (\ref{h_phi}), from (\ref{V_dash_phi_}) and (\ref{V_doubledash_phi_}) we, respectively, get
\begin{eqnarray}\label{eta_phi_0}
\eta(\phi) = 0,
\end{eqnarray}
\begin{eqnarray}\label{inequality_6}
\frac{d\eta}{d\phi} < 0,
\end{eqnarray}
where
\begin{eqnarray}\label{eta_phi}
\eta(\phi) \equiv \frac{\sqrt{2h(\phi)}}{\sqrt{g_{+}(\phi)-\phi}+\sqrt{g_{-}(\phi)-\phi}}-\frac{dS}{d\phi},
\end{eqnarray}
\begin{eqnarray}\label{g_plus_minus_phi}
g_{\pm}(\phi) = \frac{1}{2}\big(\sqrt{h(\phi)}\pm\sqrt{3\sigma_{ie}}\big)^{2}.
\end{eqnarray}
Now, the double layer solution of the energy integral (\ref{energy_integral_1}) having amplitude $ |\phi_{dl}| $ exists at $ M = M_{dl}$, where $ M_{dl}$ is given by the following equations
\begin{eqnarray}\label{M_dl}
M_{dl} = \sqrt{\frac{h(\phi_{dl})}{M_{s}}},
\end{eqnarray}
\begin{eqnarray}\label{eta_phi_dl}
\eta(\phi_{dl}) = 0,
\end{eqnarray}
with
\begin{eqnarray}\label{inequality_7}
h(\phi_{dl})-M_{c}^{2} > 0,
\end{eqnarray}
\begin{eqnarray}\label{inequality_8}
g_{-}(\phi_{dl})-\phi_{dl} \geq 0,
\end{eqnarray}
\begin{eqnarray}\label{inequality_9}
\frac{d\eta}{d\phi}\bigg|_{\phi=\phi_{dl}} < 0.
\end{eqnarray}
To derive condition (\ref{inequality_8}), we have used the following restriction  on
$ \phi $ : $ \phi \leq \Psi_{M} = \frac{1}{2}(MM_{s}-\sqrt{3\sigma_{ie}})^{2} $.
If the inequalities given by (\ref{inequality_7}), (\ref{inequality_8}) and (\ref{inequality_9}) hold simultaneously then one can get a PPDL or NPDL at $\phi=\phi_{dl}$ according to whether $\phi_{dl}>0$ or $\phi_{dl}<0$.

So, if the system supports any double layer solution, we can easily find the Mach number corresponding to that double layer solution by using the equations (\ref{M_dl}) - (\ref{eta_phi_dl}) and the conditions (\ref{inequality_7}) - (\ref{inequality_9}). Now, if there exist solitary waves after the formation of double layer then our aim is to investigate whether the occurrence of PPSWs (NPSWs) after the formation of PPDL (NPDL) are bounded or not. The amplitudes of PPSWs or NPSWs are also monotonically increasing for increasing values of Mach number after the formation of double layer. Now we want to investigate whether there exist any Mach number for which the solitary wave takes the largest amplitude after the formation of double layer. If such Mach number exists then this Mach number restricts the occurrence of PPSWs (NPSWs) after the formation of PPDL (NPDL) and consequently the existence region of PPSWs (NPSWs) becomes bounded.

Secondly, if the system does not support either PPDL or NPDL or both then our aim is to investigate whether the occurrence of PPSWs or NPSWs are bounded or not. So, in this case our task is to find the upper bound of the Mach number (if exists) for the occurrence of PPSWs or NPSWs. In the next two subsections, we shall analytically investigate the upper bounds of the Mach number for the existence of PPSWs (including positive potential supersolitons) and NPSWs (including negative potential supersolitons) of the energy integral (\ref{energy_integral_1}).

\subsection{\label{sec:supersolitons} Upper bound of the Mach number for the existence of PPSWs (including positive potential supersolitons)}

To find an upper bound of the Mach number for the existence of any type of positive potential solitary structures, following \citet{das12}, we consider the existence of a PPSW of amplitude $ \phi_{m}  > 0 $. Then, following conditions are simultaneously satisfied.
\begin{eqnarray}
&& V (0)=0, V' (0)=0 \mbox{      and      } V'' (0)<0 \label{inequality_1A}\\
&& V(\phi_{m}) = 0 \mbox{      and      } V'(\phi_{m}) > 0 \label{inequality_1B}\\
&& V(\phi) < 0    \mbox{      for all      } 0 < \phi < \phi_{m}\label{inequality_1}
\end{eqnarray}
The conditions as given in (\ref{inequality_1A}) are simultaneously satisfied if $M>M_{c}=1$. We assume that this condition holds good. We also assume that the conditions as given in (\ref{inequality_1B}) are also true. Consider the condition as given in (\ref{inequality_1}).

We have seen in the previous section that $ V(\phi) $ is real only when $ \phi \leq \Psi_{M} $ and consequently, we must have $ \phi_{m} \leq \Psi_{M} $ , otherwise $ V(\phi_{m}) $ is not a real quantity. Therefore, we can rewrite the inequality as given in (\ref{inequality_1}) as
\begin{eqnarray}\label{inequality_2}
V(\phi) < 0    \mbox{      for all      } 0 < \phi < \phi_{m} \leq  \Psi_{M}
\end{eqnarray}
But the inequality (\ref{inequality_2}) can define a large amplitude PPSW of amplitude $\Psi_{M}$, which is in conformity with (\ref{inequality_1A}) if
\begin{eqnarray}\label{inequality_2A}
V(\Psi_{M}) = 0 \mbox{      and      } V'(\Psi_{M}) > 0 .
\end{eqnarray}
Again, let $ M_{max} $  be the maximum value of $ M $ up to which positive potential solitary wave solution can exist. As $ \Psi_{M} $  increases with $ M $, $ \Psi_{M} \leq \Psi_{M_{max}} $. Therefore, the inequality
\begin{eqnarray}\label{inequality_3}
V(\phi) < 0    \mbox{      for all      } 0 < \phi < (\phi_{m} \leq  \Psi_{M} \leq ) \Psi_{M_{max}}
\end{eqnarray}
defines the largest amplitude PPSW if
\begin{eqnarray}\label{inequality_3A}
V(\Psi_{M_{max}}) = 0 \mbox{      and      } V'(\Psi_{M_{max}}) > 0 .
\end{eqnarray}
Therefore, for the existence of PPSWs, the  Mach  number $ M $ is  restricted  by  the  following inequality: $ M_{c} < M \leq M_{max} $, where $ M_{max} $ is the largest positive root of equation $ V(\Psi_{M}) = 0 $ subject to the condition $ V(\Psi_{M}) \geq 0 $ for all $ M \leq M_{max} $.

So, $M_{max}$ can restrict the existence of PPSWs. But if a PPDL exists at $ M = M_{PPDL} $, then this PPDL can restrict the existence of at least one sequence of PPSWs because the existence of a PPDL implies the existence of at least one sequence of PPSWs which converges to the PPDL solution. Therefore, with respect to the existence of $M_{PPDL}$ and $M_{max}$, we have the following three cases.

\textbf{Case-1 :} If $M_{PPDL}$ exists but $M_{max}$ does not exist for fixed values of the parameters involved in the system then obviously $M_{PPDL}$ is the upper bound of $M$ for the existence of positive potential solitary waves, i.e., one can get a PPSW for all $M$ such that $M_{c}<M<M_{PPDL}$ and for $M>M_{PPDL}$ there does not exist any positive potential solitary structure whereas at $M=M_{PPDL}$ one can get a PPDL solution.

\textbf{Case-2 :} If $M_{max}$ exists but $M_{PPDL}$ does not exist for fixed values of the parameters involved in the system then $M_{max}$ is the upper bound of $M$ for the existence of positive potential solitary waves, i.e., one can get a PPSW for all $M$ such that $M_{c}<M \leq M_{max}$ and for $M>M_{max}$ there does not exist any positive potential solitary structure.

\textbf{Case-3 :} If both $M_{max}$ and $M_{PPDL}$ exist for fixed values of the parameters involved in the system then the case for which $M_{c}<M_{max}<M_{PPDL}$ is not possible for the following reasons: The existence of PPSW is restricted by the inequality $M_{c}<M<M_{PPDL}$ only when there does not exist $M_{max}$. In other words, existence of $M_{max}$ implies that the mach number $M$ for existence of any type of positive potential solitary structure is always restricted by the inequality $M_{c}<M \leq M_{max}$. In this case, we have seen that $M_{max}$ exists finitely but $M_{max}$ is not the upper bound of the mach number $M$ for the existence of PPSWs, which is not possible because there exist PPSWs for all $M$ lying within the interval $M_{max}<M<M_{PPDL}$. So, if $M_{max}$ exists finitely, then $M_{max}$ must be the upper bound of the mach number $M$ for the existence of any type of PPSWs, i.e., there does not exist any positive potential solitary structures for $M>M_{max}$.

Therefore, if both $M_{max}$ and $M_{PPDL}$ exist for fixed values of the parameters involved in the system then we must have $M_{c}<M_{PPDL} < M_{max}$. For this case we have the following conclusions: For this case, we can split the entire range of the Mach number $M$ into two disjoint subintervals, viz., $M_{c}<M<M_{PPDL}$ and $M_{PPDL}<M \leq M_{max}$. Now, for $M_{c}<M<M_{PPDL}$, we get a sequence of PPSWs converging to the positive potential double layer solution at $M=M_{PPDL}$. In other words, the positive potential double layer solution at $M=M_{PPDL}$ can restrict the occurrence of all PPSWs for all $M$ lying within the interval $M_{c}<M<M_{PPDL}$  whereas for $M_{PPDL}<M \leq M_{max}$, we get positive potential solitons after the formation of positive potential double layer at $M=M_{PPDL}$, i.e., we get dias type solitons (according to \citet{das12}) or supersolitons (according to \citet{dubinov12a}), which has been recently reported by many authors for some particular values of the parameters involved in the system. For $M_{PPDL}<M \leq M_{max}$, amplitude of the supersoliton increases with increasing values of $M$ but the amplitude of the supersolitons are bounded and the maximum amplitude of the supersoliton can be obtained at $M=M_{max}$. So, we have two types of PPSWs: the first type is restricted by $M_{c}<M<M_{PPDL}$ whereas the second type is restricted by $M_{PPDL}<M \leq M_{max}$. In fact, \citet{das12} clearly stated the following conditions for existence of supersolitons or dias type solitons. Supersolitons can exist if there exist two types of solitary waves of same polarity separated by a double layer of same polarity and in this case there exists a jump type discontinuity between the amplitudes of the solitary waves separated by a double layer.  In their paper, they have also reported that if there exists two types of NPSWs (PPSWs) separated by a NPDL (PPDL) then there is a finite jump between the amplitudes of two types of NPSWs (PPSWs) only when $\partial V /\partial M < 0$ for all $M > 0$ and all $ \phi < 0 (\phi > 0)$.

On the other hand, if there exists a PPDL solution $\phi = \phi_{PPDL} (>0)$ of the energy integral (\ref{energy_integral_1}) for the Mach number $M=M_{PPDL}$, then $\phi=\phi_{PPDL}$ is the smallest positive double root of the equation $V(\phi)=0 (\equiv V(M_{PPDL},\phi)=0)$ such that $0< \phi_{PPDL} \leq \Psi_{M_{PPDL}}$ i.e., $V(M_{PPDL}, \phi_{PPDL})=0$ and $V'(M_{PPDL}, \phi_{PPDL})=0$ along with the condition $0<\phi_{PPDL} \leq \Psi_{M_{PPDL}}$. If $\phi=\phi_{PPDL} (>0)$ is the only root (double root) of the equation $V(\phi) (\equiv V(M_{PPDL},\phi)=0)$ then this PPDL solution is the ultimate solution of the energy integral (\ref{energy_integral_1}) and in this case, it is not possible to get any other PPSWs for $M > M_{PPDL}$, i.e., for the occurrence of any PPSW, the Mach number $M$ is restricted by the inequality $M_{c}<M<M_{PPDL}$. But if there exists a simple positive root $\phi_{D1}(>0)$ of $V(M_{PPDL},\phi)=0$ for the unknown $\phi$ just after the occurrence of the double layer solution at $\phi = \phi_{PPDL} (>0)$ for the Mach number $M=M_{PPDL}$, i.e., if $V(M_{PPDL},\phi_{D1})=0$ along with the conditions $V(M_{PPDL},\phi_{PPDL})=0$, $V'(M_{PPDL},\phi_{PPDL})=0$, $0< \phi_{PPDL}< \phi_{D1} \leq \Psi_{M_{PPDL}}$ and $V(M_{PPDL},\phi)<0$ for all $\phi_{PPDL} < \phi < \phi_{D1}$, then there exists a PPSW solution of the energy integral (\ref{energy_integral_1}) for at least one value of $M > M_{PPDL}$, and consequently, we get a PPSW solution of the energy integral (\ref{energy_integral_1}) after the formation of PPDL at $M=M_{PPDL}$, i.e., there exists positive potential supersoliton. Although the amplitudes of supersolitons are monotonically increasing with increasing values of $M$ for $M > M_{PPDL}$, but the amplitudes of the supersolitons are restricted by $\Psi_{M}$ since $0<\phi \leq \Psi_{M}$. Therefore, one can find a Mach number $M$ for which the amplitude of the supersolitons attains its maximum value and from the previous discussion, we have seen that the amplitude of the supersolitons attains its maximum value when $M=M_{max}(>M_{PPDL})$.

\subsection{Upper bounds of the Mach number for the existence of NPSWs (including negative potential supersolitons)}

We have seen earlier that $ V(\phi) $ is real if $ \phi \leq \Psi_{M} $ , where $ \Psi_{M} $  is strictly positive. For NPSWs or NPDLs, we have
\begin{eqnarray}\label{inequality_4}
V(\phi) < 0    \mbox{      for all      } \phi_{m} < \phi < 0
\end{eqnarray}
along  with  the  conditions  stated  in  Section \ref{sec:energy_int}  for  the existence  of  NPSWs  or  NPDLs.  As  $ \Psi_{M} $  is  strictly positive and for NPSWs or NPDLs
$ \phi < 0 $, the condition $ \phi < \Psi_{M} $ is automatically satisfied and consequently for these  two  cases (NPSWs and NPDLs)  $ V(\phi) $  is  well  defined  for  all
$ \phi < 0 $ without imposing any extra condition.  Since there is no such restriction on $ \phi $, we cannot use the same definition as in the case of PPSWs to find the upper bound of Mach numbers for the existence of NPSWs. For the case of NPSWs, to find an upper limit or upper bound of $ M $ up to which NPSW can exist, we shall first find a value $ M_{NPDL} $  of $ M $ for which energy integral (\ref{energy_integral_1}) gives a NPDL solution at $ M = M_{NPDL} $ with amplitude
$ \phi = \phi_{NPDL} $.  Now, if at $ M = M_{NPDL} $, $ \phi = \phi_{NPDL} $  is the only root (double root) of equation $ V(\phi) \equiv V(M, \phi) = 0 $, i.e.
$ V(M_{NPDL}, \phi_{NPDL}) = 0 $ and $ V'(M_{NPDL}, \phi_{NPDL}) = 0 $, then the NPDL solution is the ultimate solution of the energy integral (\ref{energy_integral_1}), and in this case no NPSW solution can be obtained for $ M > M_{NPDL} $, i.e. for the occurrence of NPSWs the Mach number $M$ is restricted by the inequality
$ M_{c} < M < M_{NPDL} $.

On  the  other  hand,  if  there  exists  an  inaccessible simple negative root $ \phi_{D2}(<0) $  of $ V(M_{NPDL},\phi)= 0 $ for the unknown $ \phi $ such that $ |\phi_{D2}| > | \phi_{NPDL}| $ , i.e., if $ V(M_{NPDL},\phi_{D2}) = 0 $ along with the conditions $ V(M_{NPDL},\phi_{NPDL}) = 0 $, $ V'(M_{NPDL}$ $,\phi_{NPDL}) = 0 $, $ |\phi_{D2} | >  |\phi_{NPDL}| $, and $ V(M_{NPDL},\phi) < 0 $ for all $  \phi_{D2}  < \phi < \phi_{NPDL} $,   then there exists a NPSW solution of the energy integral (\ref{energy_integral_1}) for at least one value of $ M > M_{NPDL} $. Therefore, the double layer solution is unable to restrict the occurrence of NPSWs or there exist NPSWs for all $ M > M_{NPDL} $, i.e., in this case, there exist negative potential unbounded supersolitons. For example, one can consider the DIA solitary structures in the plasma system considered by \citet{das12}.

On the basis of the analytical theory, which have been considered in the present section, we have developed a computational scheme to draw the solution space or the compositional parameter space showing the nature of existence of different solitary structures of the present system. More specifically, a solution space or compositional parameter space is a figure where we plot the curves $M=M_{c}$, $M=M_{max}$, $M=M_{PPDL}$ and $M=M_{NPDL}$ with respect to any parameter of the system and consequently, in the compositional parameter space, if we can find a region where the existence region of PPSWs (NPSWs) is separated by the curve $M=M_{PPDL}$ ($M=M_{NPDL}$), then we can claim that the system supports positive (negative) potential supersolitons. In particular, for the case of PPSWs, if we can find a region where $M_{PPDL}<M_{max}$ then we can claim that the system supports positive potential supersolitons for all Mach number $M$ lying within the interval $M_{PPDL}<M \leq M_{max}$. So, if we can draw the solution space then it is simple to find the existence region of supersolitons and we can avoid the `trial and error' method to investigate the existence of supersolitons for particular values of the parameters involved in the system. In the next section, we have investigated different solitary structures associated with the different solutions of the energy integral (\ref{energy_integral_1}) with the help of the qualitatively different compositional parameter spaces.

\section{\label{sec:solution_spaces} Different solution spaces of the energy integral}

Four basic parameters of the present e-p-i-d plasma system are $ p$,  $\mu$,  $\sigma_{pe}$ and $\sigma_{ie}$, which are, respectively, the ratio of unperturbed number density of  positrons to that of the total unperturbed number density of positive charge, the ratio of unperturbed number density of electrons to that of the total unperturbed number density of negative charge, the ratio of the average temperature of positrons to that of electrons and the ratio of the average temperature of ions to that of electrons. As the parameter $\sigma_{ie}$ assumes a constant value for dusty plasma system, we will discuss the existence regions of the different solitary structures with respect to $\mu$ for different values of $p$ and $\sigma_{pe}$ and with respect to $p$ for different values of $\mu$ and $\sigma_{pe}$. A solution space or compositional parameter space is a figure where we plot the curves $M=M_{c}$, $M=M_{max}$, $M=M_{PPDL}$ and $M=M_{NPDL}$ with respect to any parameter of the system. In the present paper the solution spaces have been drawn with respect to the parameter $p$  or $\mu$ . To interpret the compositional parameter space we have made a general description as follows:
\begin{enumerate}
  \item Solitary structures start to exist just above the lower curve $ M = M_{c} =1 $.
  \item At each point on the curve $M=M_{PPDL}$ ($M=M_{NPDL}$), one can get a PPDL (NPDL) solution.
  \item In absence of $M_{PPDL}$ ($ M_{max}$), $ M_{max}$ ($ M_{PPDL}$) is the upper bound of $ M $ for the existence of PPSWs i.e there does not exist any PPSW if $ M > M_{max}$ ($ M> M_{PPDL}$). Although, it is important to note that there exist PPSWs along the curve $M=M_{max}$ but there does not exist any PPSW along the curve $M=M_{PPDL}$.
  \item If both $M_{PPDL}$ and $M_{max}$ exist finitely, then $\max\{M_{max},M_{PPDL}\}$ is the upper bound of $ M $ for the existence of PPSWs. The case has already been discussed in subsection \ref{sec:supersolitons} of section \ref{sec:L_U_Bounds}.
  \item If we pick a $p$ or $\mu$ and go vertically upwards, then all intermediate values of $ M $ bounded by the curves $ M=M_{c} $ and $ M=M_{max} $ or $M_{PPDL}$ or $\max\{M_{max},M_{PPDL}\}$ would give PPSWs.
  \item Similarly, all intermediate values of $ M $ bounded by the curves $ M=M_{c} $ and $ M=M_{NPDL} $ would give NPSWs.
  \item If both $M_{PPDL}$ and $M_{max}$ exist and $M_{PPDL}<M_{max}$ then all intermediate values of $ M $ bounded by the curves $ M=M_{PPDL} $ and $ M=M_{max} $ would give positive potential supersolitons. In fact, for $M_{c}<M<M_{PPDL}$, one can get a sequence of PPSWs converging to the positive potential double layer solution at $M=M_{PPDL}$ whereas for $M_{PPDL}<M \leq M_{max}$, one can get positive potential supersolitons. For $M_{PPDL}<M \leq M_{max}$, amplitude of the supersoliton increases with increasing values of $M$ but for the present system the amplitudes of the supersolitons are bounded and the maximum amplitude of the supersoliton can be obtained at $M=M_{max}$. As a result, we get two types of PPSWs separated by a PPDL and the existence of first type PPSWs are restricted by $M_{c}<M<M_{PPDL}$ whereas the existence of second type PPSWs (i.e., positive potential supersolitons) are restricted by $M_{PPDL}<M \leq M_{max}$. A finite jump between the amplitudes of two types of PPSWs separated by a PPDL can be easily verified by plotting $V(\phi)$ against $\phi$ or by plotting $\phi$ against $\xi$.
  \item We have used the following notations to interpret the different solution spaces  : C -- Region of coexistence of both NPSWs and PPSWs, N -- Region of existence of NPSWs, P -- Region of existence of PPSWs and S -- Region of existence of Supersolitons.
\end{enumerate}

\noindent Now we consider the compositional parameter spaces or solution spaces with respect to the parameter $\mu$ for increasing values of $p$ starting from $p=0$, i.e., there is no positron in the system.

\noindent \textbf{Solution Space w.r.t $\mu$ when $p=0$ :} Figure \ref{sol_spc_wrt_mu_p=0} shows the compositional parameter space or solution space with respect to the parameter $\mu$ for $p=0$, i.e., there is no positron in the system. From this figure, we have the following observations:
\begin{itemize}
    \item The system supports NPSW for all $M>M_{c}$.
    \item The system does not support any NPDL.
    \item The system does not support any PPDL.
    \item According to the definition of supersolitons, the system does not support supersoliton of any polarity because the system does not support double layer solution of any polarity.
    \item PPSWs start to exist if $\mu$ exceeds a critical value $\mu_{c}$ and for all $M$ lying within $M_{c}<M \leq M_{max}$.
    \item The system supports coexistence of both NPSWs and PPSWs if $\mu>\mu_{c}$ and $M_{c}<M \leq M_{max}$. For $\sigma_{ie}=\sigma_{pe}=0.9$, the value of $\mu_{c}$ is 0.145.
\end{itemize}

\begin{figure}
  \includegraphics{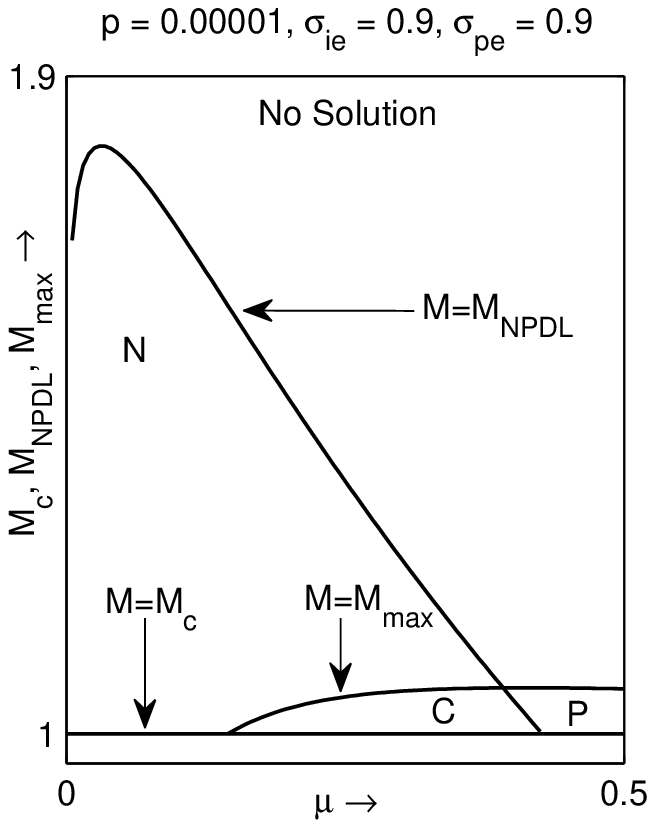}
  \caption{\label{sol_spc_wrt_mu_p=0_pt_00001} Compositional parameter space with respect to $\mu$ for $p=0.00001$, $\sigma_{ie}=0.9$ and $\sigma_{pe}=0.9$}.
\end{figure}

\noindent \textbf{Solution Space w.r.t $\mu$ for $0<p<0.0001$:} Figure \ref{sol_spc_wrt_mu_p=0_pt_00001} shows the compositional parameter space or solution space with respect to the parameter $\mu$ for $p=0.00001$, i.e., a very small amount of positron is injected in the system. Here, the main important and interesting differences from the solution space with $p=0$ are as follows:
      \begin{itemize}
        \item The system supports NPDL along the curve $M=M_{NPDL}$.
        \item NPSWs are restricted for $M_{c}<M<M_{NPDL}$, i.e., the existence region of NPSWs is bounded by the curves $M=M_{c}$ and $M=M_{NPDL}$ and consequently, the system does not support any negative potential supersoliton.
        \item The region of coexistence of NPSWs and PPSWs is bounded by the curves $M=M_{c}$, $M=M_{max}$ and $M=M_{NPDL}$.
        \item The system supports coexistence of NPDL and PPSW along the curve $M=M_{NPDL}$ bounded by the curves $M=M_{c}$ and $M=M_{max}$.
        \item As the system does not support any PPDL solution, the system does not support any positive potential supersoliton.
      \end{itemize}
For increasing values of $p$ lying within $0<p<0.0001$, the qualitative behaviour of the solution spaces remains unchanged, the only exception is that the region of existence of NPSWs decreases with increasing values of $p$ for $0<p<0.0001$.

\noindent \textbf{Solution Space w.r.t $\mu$ for $0.0001 \leq p <0.025$ :} If we further increase the concentration of positron from $p=0.0001$ to $p<0.025$, the solution space is not qualitatively same as solution space for $0<p<0.0001$. The main difference in this solution space is the existence of PPDL solution along the curve $M=M_{PPDL}$ for a very small interval of $\mu$ which has been shown in figure \ref{sol_spc_wrt_mu_p=0_pt_01} for $p=0.01$. Again, it has also been observed that as $p$ increases within $0.0001 \leq p <0.025$, the interval of $\mu$ for the existence of PPDL solution increases. Again, the existence of PPDL solution implies the existence of PPSWs bounded by the curves $M=M_{c}$ and $M=M_{PPDL}$. From the solution space as given in figure \ref{sol_spc_wrt_mu_p=0_pt_01}, we can again conclude that the region of existence of NPSWs decreases with increasing values of $p$ for $0.0001 \leq p <0.025$. This fact can be confirmed with the help of the solution space for $p=0.02$ where we have seen a huge reduction in the existence region of NPSWs. The solution space with respect to the parameter $\mu$ is given in figure \ref{sol_spc_wrt_mu_p=0_pt_02} for $p=0.02$. It is observed that the existence region of NPSWs ultimately collapses for $p=0.025$. In fact, the existence region of NPSWs decreases for increasing values of positron concentration and finally, there exists a critical value of $p$ for which the system does not support any negative potential solitary structure (NPSW or NPDL). This fact can be efficiently described by the figure \ref{NPDLs_for_different_p}. The figure \ref{NPDLs_for_different_p} is the solution space with respect to $\mu$ for different values of $p$ showing the existence regions of NPSWs bounded by the curves $M=M_{c}$ and $M=M_{NPDL}$. But in case of the the existence region of PPSWs bounded by the curves $M=M_{c}$ and $M=M_{PPDL}$, we have seen that the existence region of PPDLs increases for increasing values of positron concentration $p$ if $p< p^{(c)}$, where $p^{(c)}$ is a certain critical value of $p$ whereas for $p>p^{(c)}$, the existence region of PPDLs decreases for increasing values of positron concentration $p$ and finally, there exists a critical value of $p$ for which the system does not support any positive potential double layer. This fact can be efficiently described by the figure \ref{PPDLs_for_different_p}. The figure \ref{PPDLs_for_different_p} is the solution space with respect to $\mu$ for different values of $p$ showing the existence regions of PPDLs. On the other hand, it can be easily verified that the existence region of PPSWs bounded by the curves $M=M_{c}$ and $M=M_{max}$ increases for increasing values of positron concentration $p$. For $\sigma_{ie}=\sigma_{pe}=0.9$, the value of $p^{(c)}$ is 0.0215. Again from the solution spaces as given in figure \ref{sol_spc_wrt_mu_p=0_pt_01} and figure \ref{sol_spc_wrt_mu_p=0_pt_02}, it is evident that the system does not support any positive or negative potential supersoliton.

\noindent \textbf{Solution Space w.r.t $\mu$ for $0.025 \leq p <0.07$ :} If we further increase the concentration of positron from $p=0.025$ to $p<0.07$, then the solution space is not qualitatively same as solution space for $0.0001 \leq p <0.025$. For $p=0.03$, the solution space with respect to the parameter $\mu$ is given in figure \ref{sol_spc_wrt_mu_p=0_pt_03}. Here we have the following observations:
      \begin{itemize}
        \item The system does not support any NPSWs.
        \item The existence region of PPSWs has increased.
        \item There exist two cutoff values $\mu_{a}$ and $\mu_{b}$ such that for $\mu_{a}<\mu<\mu_{b}$, we have $M_{PPDL}<M_{max}$, i.e., in this interval of $\mu$, there exist PPSWs after the formation of double layers if the mach number $M$ is restricted by $M_{PPDL}<M \leq M_{max}$. In fact, we have two types of PPSWs: First Type - bounded by the curves $M=M_{c}$ and $M=M_{PPDL}$ and Second Type - bounded by the curves $M=M_{PPDL}$ and $M=M_{max}$. Therefore, if these two types of PPSWs are separated by a PPDL, i.e., if the existence region or a portion of existence region of the total population of PPSWs (i.e., union of First and Second Type PPSWs) is separated by the curve $M=M_{PPDL}$  or a portion of the curve $M=M_{PPDL}$, then for this particular portion of the curve $M=M_{PPDL}$, we have a positive potential solitary wave after the formation of positive potential double layer. According to the property as mentioned by \citet{das12}, we have a jump type of discontinuity between the amplitudes of these two types of solitary waves just before and just after the formation of double layer and consequently, formation of positive potential supersoliton is confirmed. Now, for the present problem, in the existence region of supersolitons, we have $M_{PPDL}<M \leq M_{max}$ as supersoliton starts to exist just after the formation of PPDL and consequently, we have the bounded positive potential supersolitons. So, the region of existence of supersolitons is bounded by the curves $\mu=\mu_{a}$, $\mu=\mu_{b}$, $M=M_{PPDL}$ and $M=M_{max}$. It is not difficult to find the values of $\mu_{a}$ and $\mu_{b}$. For $\sigma_{ie}=\sigma_{pe}=0.9$, $p=0.03$, the values of $\mu_{a}$ and $\mu_{b}$ are 0.069 and 0.081 respectively.
        \item In Figure \ref{profile_supersliton}, the existence of positive potential supersolitons has been verified by plotting $V(\phi)$ against $\phi$ and also by plotting $\phi$ against $\xi$ for a fixed $\mu$ lying within $\mu_{a}<\mu<\mu_{b}$ and $M=M_{PPDL}+0.0001$ along with the other values of the parameters as shown in the figure. In fact, we can take any $\mu $ lying within $\mu_{a}<\mu<\mu_{b}$ and any $M$ lying within $M_{PPDL}<M \leq M_{max}$ to verify the the existence of positive potential supersolitons.
        \item It is observed that the existence region of supersolitons ultimately collapses. In fact, the existence region of supersolitons decreases for increasing values of positron concentration and finally, there exists a critical value of $p$ for which the system does not support any positive potential supersoliton. This fact can be efficiently described by the figure \ref{supersolitons_different_p}.
      \end{itemize}
From these observations, we can conclude that there must exist a critical value of $p$ where the NPSW collapses and there must exist a critical value of $p$ for which there exist two cutoff values $\mu_{a}$ and $\mu_{b}$ such that for $\mu_{a}<\mu<\mu_{b}$, we have $M_{PPDL}<M_{max}$.

\noindent \textbf{Solution Space w.r.t $\mu$ for $0.07 \leq p < 0.117$ :} If we further increase the concentration  of positron from $p=0.07$  to $p<0.117$, then the supersolitons disappear from the system and the existence region of PPDL is still decreasing. In this case, the system supports only PPSWs for all $\mu >0$. But there exists a cut off value $\mu_{d}$ of $\mu$ such that the PPSWs are bounded by the curves $M=M_{c}$ and $M=M_{PPDL}$ for $0<\mu<\mu_{d}$ whereas for $\mu_{d}\leq \mu \leq \mu_{T}$, the PPSWs are bounded by the curves $M=M_{c}$ and $M=M_{max}$, where $\mu_{T}$ is the physically admissible upper bound of $\mu$. Here the solution space with respect to the parameter $\mu$ for $p=0.07$ is given in figure \ref{sol_spc_wrt_mu_p=0_pt_07}. For the given values of $\sigma_{ie}=0.9$ and $\sigma_{pe}=0.9$, the value of $\mu_{d}$ is 0.018.

\begin{figure}
  \includegraphics{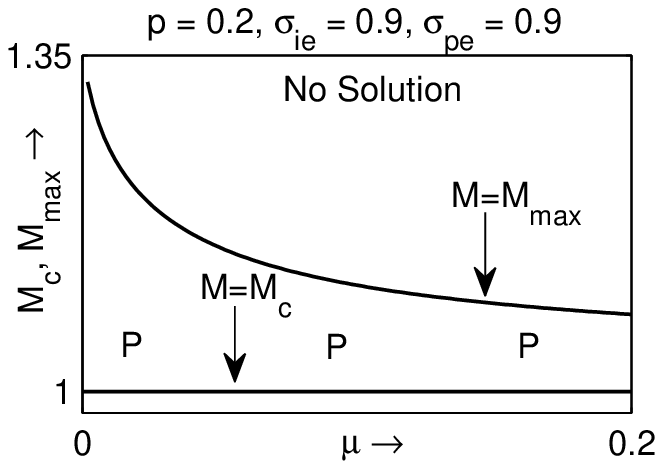}
  \caption{\label{sol_spc_wrt_mu_p=0_pt_2} Compositional parameter space with respect to $\mu$ for $p=0.2$, $\sigma_{ie}=0.9$ and $\sigma_{pe}=0.9$}
\end{figure}

\noindent \textbf{Solution Space w.r.t $\mu$ for $p \geq 0.117$ :} If we further increase the concentration of positron starting from $p = 0.117$, then the PPDL disappears from the system but the system supports only PPSWs for all $\mu >0$ and these PPSWs are bounded by the curves $M=M_{c}$ and $M=M_{max}$.  For $p=0.2$, the solution space with respect to the parameter $\mu$ is given in figure \ref{sol_spc_wrt_mu_p=0_pt_2}. This solution space is not qualitatively same as solution space for $p=0.07$. Here we see that the PPDLs disappear from the system but PPSWs exist for the entire range of $\mu$ and the existence region of PPSWs is bounded by the curves $M=M_{c}$ and $M=M_{max}$. If we further increase the concentration of positron then there is no qualitative change in the solution space, i.e., we get the similar type of solution space as obtained in case of $p=0.2$ but in any case the PPSWs are restricted by $M_{c}<M \leq M_{max}$ and $M_{max}$ increases with increasing values of $p$, i.e., the region of existence of PPSWs restricted by $M_{c}<M \leq M_{max}$ increases with increasing values of $p$ starting from $p=0.117$ and for the entire range of $\mu$.

\begin{figure}
  \includegraphics[width=8cm,height=8cm]{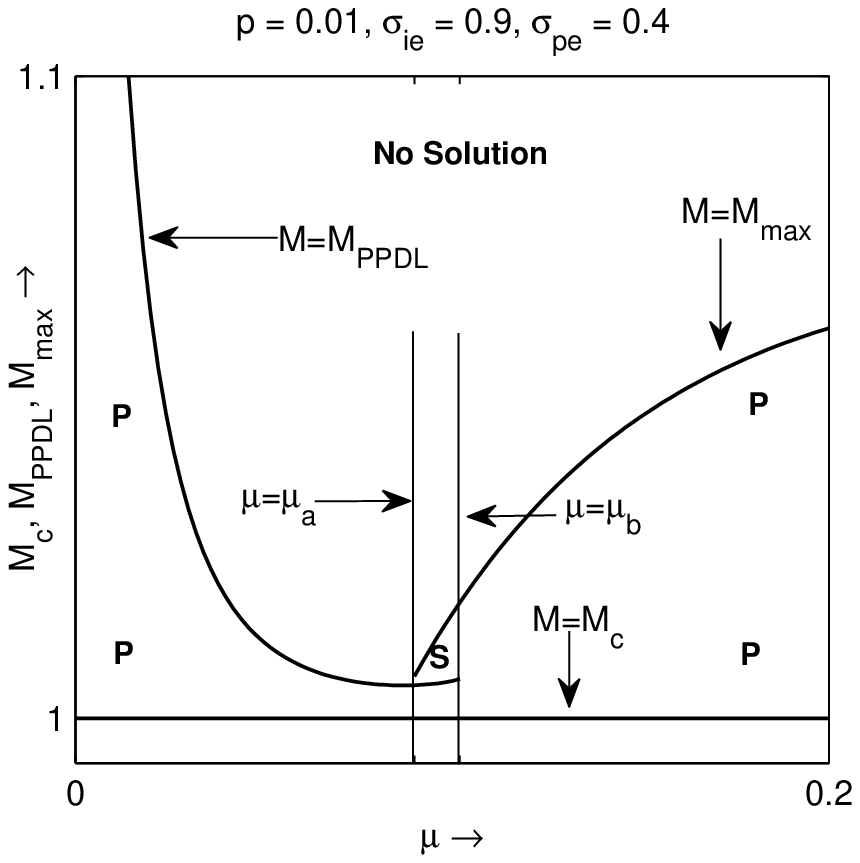}
  \caption{\label{sol_spc_wrt_mu_sigma_pe=0_pt_4} Compositional parameter space with respect to $\mu$ for $p=0.01$, $\sigma_{ie}=0.9$ and $\sigma_{pe}=0.4$}
\end{figure}

\begin{figure}
  \includegraphics[width=8cm,height=8cm]{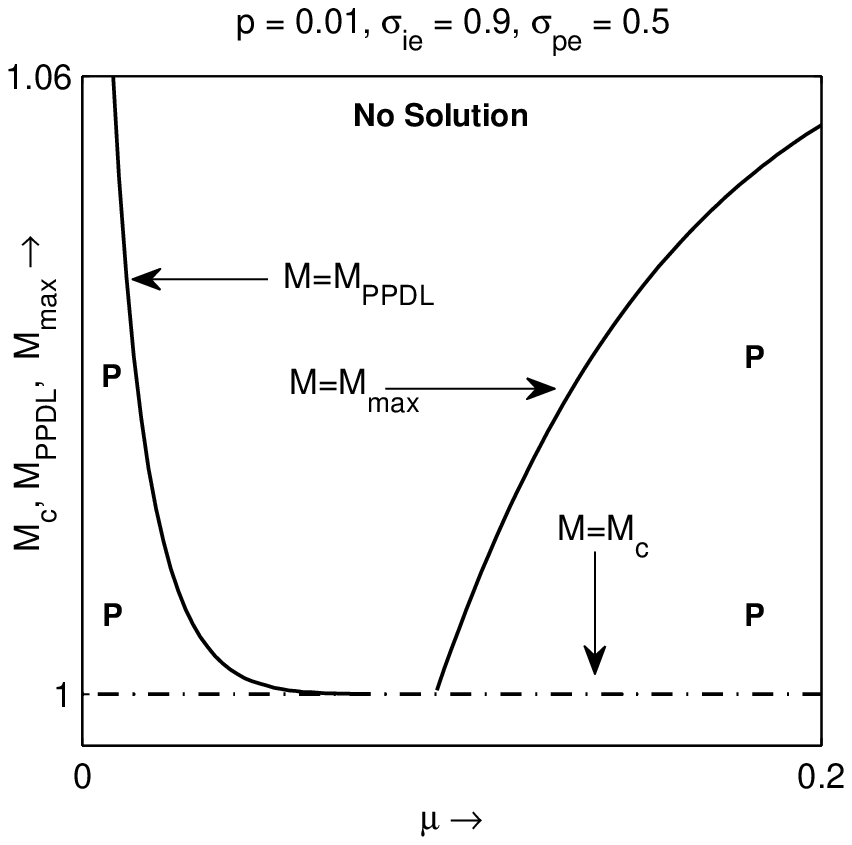}
  \caption{\label{sol_spc_wrt_mu_sigma_pe=0_pt_5} Compositional parameter space with respect to $\mu$ for $p=0.01$, $\sigma_{ie}=0.9$ and $\sigma_{pe}=0.5$}
\end{figure}

\begin{figure}
  \includegraphics[width=8cm,height=8cm]{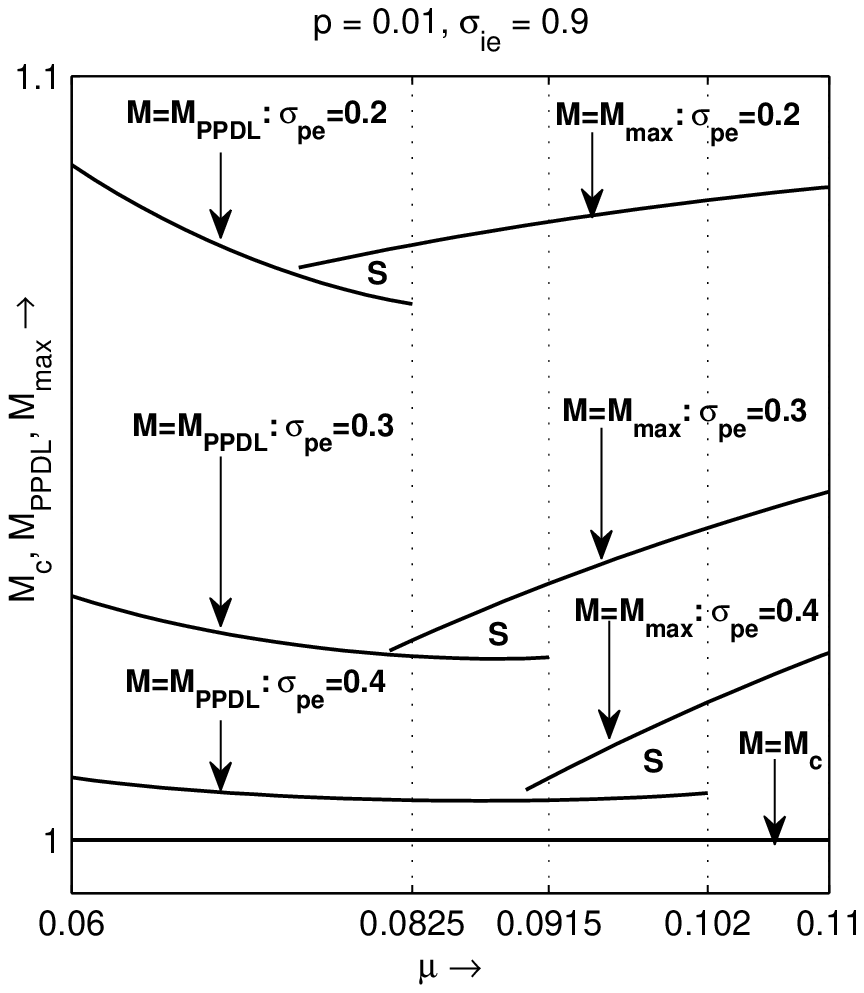}
  \caption{\label{supersolitons_diff_sigma_pe} Compositional parameter space with respect to $\mu$ showing the existence region of supersolitons for different values of $\sigma_{pe}$}
\end{figure}

\begin{figure}
  \includegraphics[width=8cm,height=8cm]{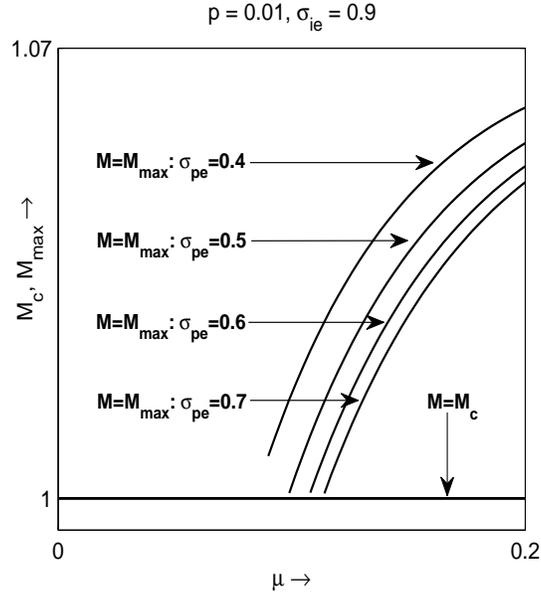}
  \caption{\label{Mmax_different_sigma_pe} Compositional parameter space with respect to $\mu$ showing the positive potential solitary waves bounded by $M=M_{max}$ for different values of $\sigma_{pe}$}
\end{figure}

\begin{figure}
  \includegraphics[width=8cm,height=8cm]{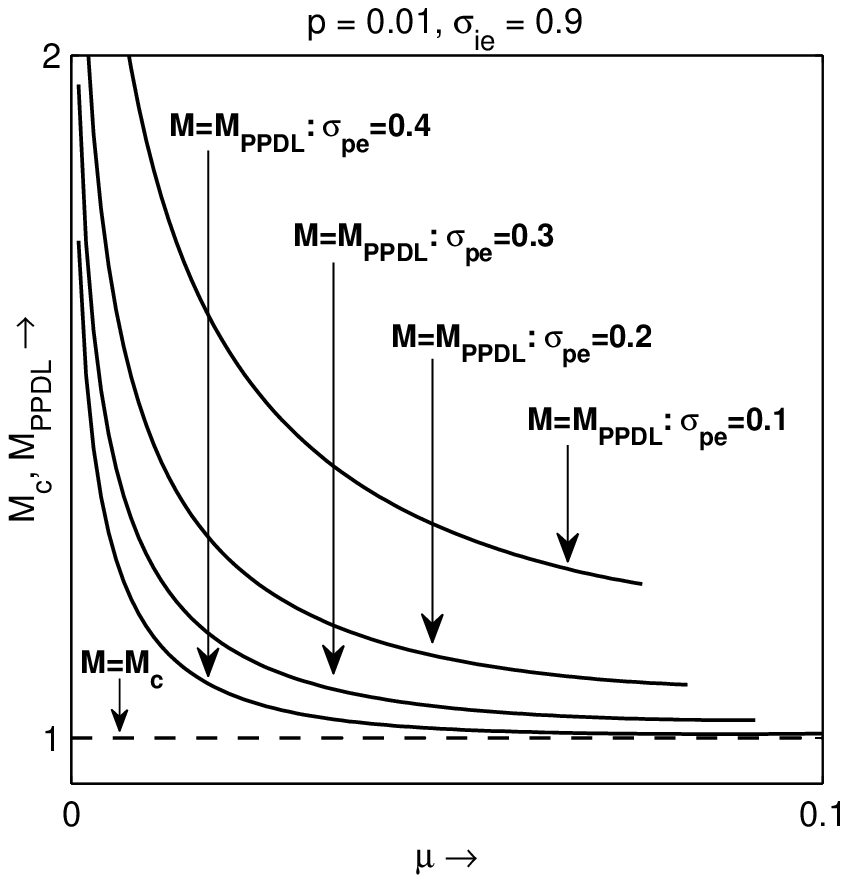}
  \caption{\label{PPDLs_different_sigma_pe} Compositional parameter space with respect to $\mu$ showing the positive potential double layers and positive potential solitary waves restricted by positive potential double layers for different values of $\sigma_{pe}$}
\end{figure}

Next we shall consider the solution spaces with respect to $\mu$ for some fixed value of $p$ but for different values of $\sigma_{pe}$.
\begin{itemize}
  \item For $p=0.01$ and $\sigma_{pe}=0.9$, we have seen in figure \ref{sol_spc_wrt_mu_p=0_pt_01} that the system does not support any supersoliton but if the value of $\sigma_{pe}$ decreases then we have seen that NPDL disappears from the system and the positive potential supersoliton comes into the picture, i.e., there exists a critical value $\sigma_{pe}^{(c)}$ of $\sigma_{pe}$ such that the system does not support any supersoliton if $\sigma_{pe}$ exceeds its critical value $\sigma_{pe}^{(c)}$ and there exists a critical value $\sigma_{pe}^{(a)}$ of $\sigma_{pe}$ such that NPDL solutions come into the picture if $\sigma_{pe} > \sigma_{pe}^{(a)}$ where $\sigma_{pe}^{(a)}>\sigma_{pe}^{(c)}$. For $\sigma_{ie}=0.9$, $p=0.01$, the values of $\sigma_{pe}^{(a)}$ and $\sigma_{pe}^{(c)}$ are 0.5445 and 0.499 respectively.
  \item Figure \ref{sol_spc_wrt_mu_sigma_pe=0_pt_4} shows the solution space with respect to $\mu$ for $p=0.01$ and $\sigma_{pe}=0.4$. This solution space shows the existence of positive potential supersolitons but nonexistence of NPDLs. In fact, for the decreasing values of $\sigma_{pe}$ starting from $\sigma_{pe}=0.4$, the system will always support supersolitons but such supersoliton disappears if $\sigma_{pe}$ exceeds the critical value $\sigma_{pe}^{(c)}$.
  \item Next, figure \ref{sol_spc_wrt_mu_sigma_pe=0_pt_5} shows the solution space with respect to $\mu$ for $p=0.01$ and $\sigma_{pe}=0.5$ and we have seen that positive potential supersoliton disappears from the system. In figure \ref{supersolitons_diff_sigma_pe}, the region of existence of positive potential supersolitons has been shown for different values of $\sigma_{pe}$. From this figure, we see that the region of existence of supersolitons increases with increasing values of $\sigma_{pe}$. In fact, it has been observed that there exists a critical value $\sigma_{pe}^{(b)}$ of $\sigma_{pe}$ such that at $\sigma_{pe}=\sigma_{pe}^{(b)}$ the region of existence of supersolitons is maximum and after $\sigma_{pe}=\sigma_{pe}^{(b)}$ the region of existence of supersolitons decreases very rapidly with increasing values of $\sigma_{pe}$ and finally it disappears from the system and after the disappearance of the positive potential supersolitons, NPDLs come into the picture as we have seen earlier for $\sigma_{pe}=0.9$. For $\sigma_{ie}=0.9$, $p=0.01$, the values of $\sigma_{pe}^{(b)}$ is very close to $\sigma_{pe}^{(c)}$.
  \item Other interesting results obtained from the solution spaces are the followings:
\begin{itemize}
  \item Region of existence of PPSWs bounded by the curves $M=M_{c}$ and $M=M_{max}$ decreases with increasing values of $\sigma_{pe}$ as shown in the figure \ref{Mmax_different_sigma_pe}.
  \item Region of existence of PPSWs bounded by the curves $M=M_{c}$ and $M=M_{PPDL}$ increases with increasing values of $\sigma_{pe}$ as shown in the figure \ref{PPDLs_different_sigma_pe}.
  \item It can also be checked that the region of existence of NPSWs bounded by the curves $M=M_{c}$ and $M=M_{NPDL}$ increases with increasing values of $\sigma_{pe}$. In fact, here we have qualitatively same type of solution spaces as given in figure \ref{NPDLs_for_different_p}. The only exception is that the region of existence of NPSWs bounded by the curves $M=M_{c}$ and $M=M_{NPDL}$ increases with increasing values of $\sigma_{pe}$ instead of the fact that the region of existence of NPSWs bounded by the curves $M=M_{c}$ and $M=M_{NPDL}$ decreases with increasing values of $p$.
  \item For small value of $p$ and for very large value of $\sigma_{pe}$ ($\sigma_{pe}=100$, say), we have exactly same type of solution space as given in figure \ref{sol_spc_wrt_mu_p=0}, i.e., the solution space for p=0. This fact is consistent with the Energy integral (\ref{energy_integral}) because it is simple to check that
  \begin{eqnarray}\label{limit}
  \lim_{\sigma_{pe} \rightarrow \infty} \sigma_{pe}W_{p}=\frac{\phi}{\Phi},
  \end{eqnarray}
  and consequently the equation (\ref{V_phi}) can be written as
  \begin{eqnarray}\label{V_phi_modified}
  	W(\phi) = \frac{\Phi^{2}}{\lambda_{D}^{2}}(M_{s}^{2}-3 \sigma_{ie})\Bigg[W_{i} - \frac{\mu}{1-p} W_{e}-\frac{1-p-\mu}{1-p}W_{d}\Bigg].
	\end{eqnarray}
	Now, for small values of $p$, we can set $1-p \approx 1$ and the above expression of $W(\phi)$ assumes the following form
	\begin{eqnarray}\label{V_phi_modified_further}
  	W(\phi) = \frac{\Phi^{2}}{\lambda_{D}^{2}}(M_{s}^{2}-3 \sigma_{ie})\Bigg[W_{i} -
  	 \mu W_{e}-(1-\mu)W_{d}\Bigg].
	\end{eqnarray}
	The equation (\ref{V_phi_modified_further}) is free from the parameters associated with the positrons and consequently, for this case, there is no effect of positrons on the solitary structures of the DIA waves. On the other hand, if $\mu + p <1$, then we have the qualitatively similar type of solution space as given in figure \ref{sol_spc_wrt_mu_p=0}.
\end{itemize}
\end{itemize}

From the solution space with respect to $\mu$ for different values of $\sigma_{pe}$ (actually, for increasing values of $\sigma_{pe}$), we have seen that the NPDL comes into the picture after the disappearance of positive potential supersolitons whereas from the solution space with respect to $\mu$ for different values of $p$ (actually, for increasing values of $p$), we have seen that the positive potential supersoliton comes into the picture after the disappearance of NPDL. Therefore, for the solution space with respect to $\mu$, we have seen that it is impossible to get simultaneous occurrence of positive potential supersolitons and NPDLs for any arbitrarily fixed values of $p$ and $\sigma_{pe}$. This fact can be easily understood from the solution space with respect to $p$. Figure \ref{sol_spc_wrt_p_mu=0_pt_05} shows the solution space with respect to $p$ for fixed values of $\mu$ and $\sigma_{pe}$ as mentioned in the figure. Although this solution space shows the simultaneous occurrence of NPDLs and positive potential supersolitons but if we restrict the value of $p$ at $p=p_{k}$ (a fixed value of $p$), then from the figure it is clear that the simultaneous occurrence of NPDL and positive potential supersoliton is not possible. In other words, if we move the solution space along the curve parallel to the curve $M=M_{c}$ then from the solution space as given in figure \ref{sol_spc_wrt_p_mu=0_pt_05}, we see that the positive potential supersoliton comes into the picture after the disappearance of all NPDLs. Again, we have seen the coexistence of NPSWs and PPSWs for an interval of $p$ bounded by the curves $M=M_{c}$, $M=M_{NPDL}$ and $M=M_{max}$ only when the positive potential supersoliton disappears from the system. But this case occurs only when $\mu$ exceeds a critical value. This qualitative behaviour has been shown in the figure \ref{sol_spc_wrt_p_mu=0_pt_15}. Figure \ref{sol_spc_wrt_p_mu=0_pt_15} shows the solution space with respect to $p$ for fixed values of $\mu$ and $\sigma_{pe}$ as mentioned in the figure. From this solution space we see that positive potential supersoliton disappears from the system and at the same time coexistence of both positive and negative potential solitary waves take place for a definite interval of $p$. If we further increase the value of $\mu$ then we see that NPDLs disappear from the solution space and for the entire interval of $p$, we have only PPSWs bounded by the curves $M=M_{c}$ and $M=M_{max}$. This type of solution space is qualitatively same as the solution space with respect to $\mu$ as shown in figure  \ref{sol_spc_wrt_mu_p=0_pt_2}.

\section{\label{sec:Summary_Discussions} Summary \& Discussions }

\begin{figure}
  \includegraphics{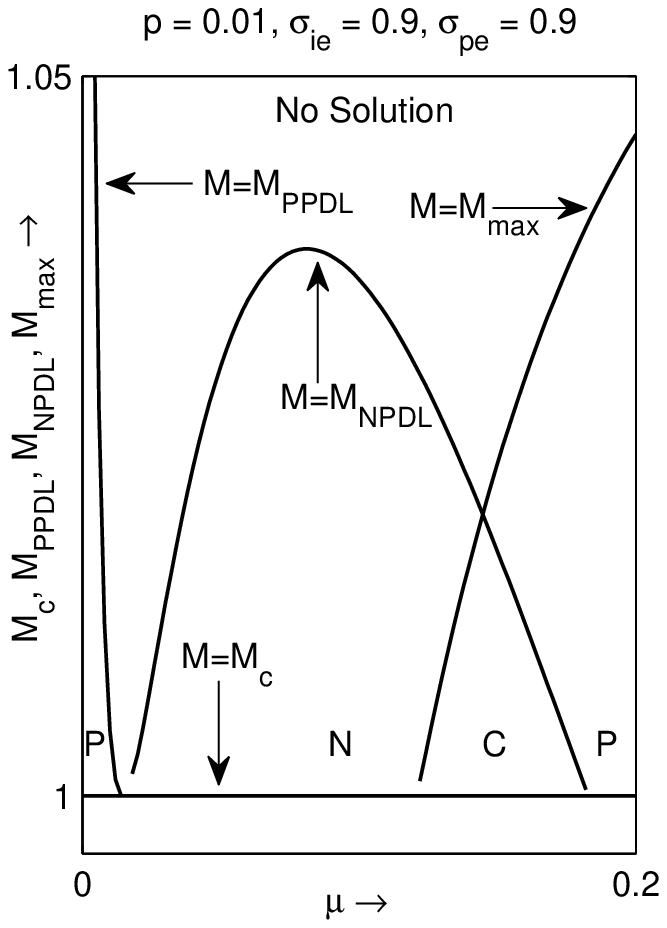}
  \caption{\label{sol_spc_wrt_mu_p=0_pt_01} Compositional parameter space with respect to $\mu$ for $p=0.01$, $\sigma_{ie}=0.9$ and $\sigma_{pe}=0.9$}.
\end{figure}

\begin{figure}
  \includegraphics{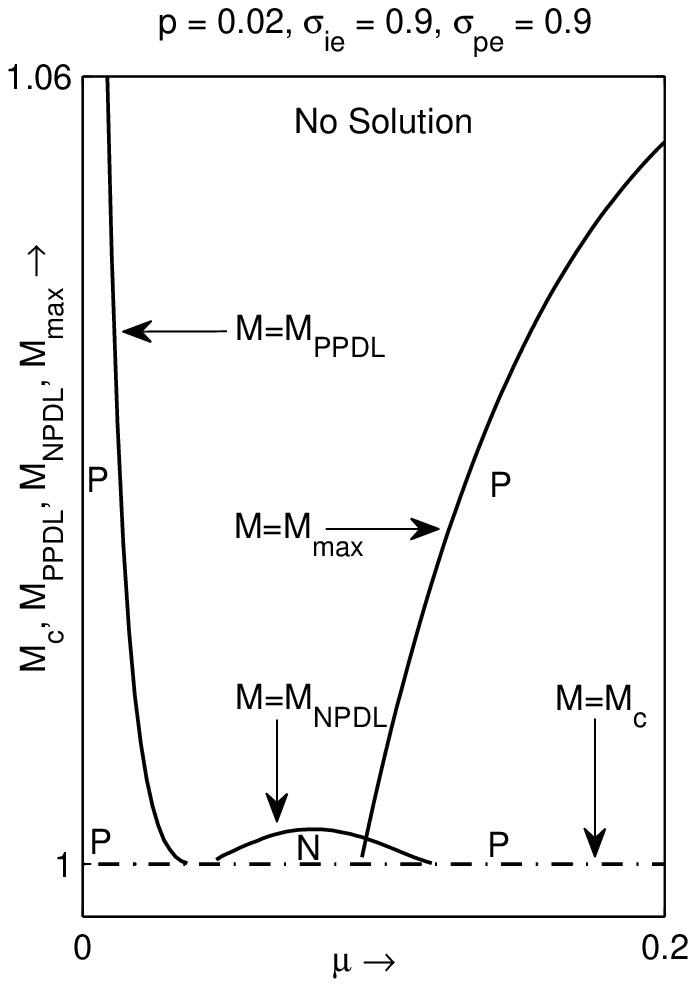}
  \caption{\label{sol_spc_wrt_mu_p=0_pt_02} Compositional parameter space with respect to $\mu$ for $p=0.02$, $\sigma_{ie}=0.9$ and $\sigma_{pe}=0.9$}.
\end{figure}

\begin{figure}
  \includegraphics[width=8cm,height=8cm]{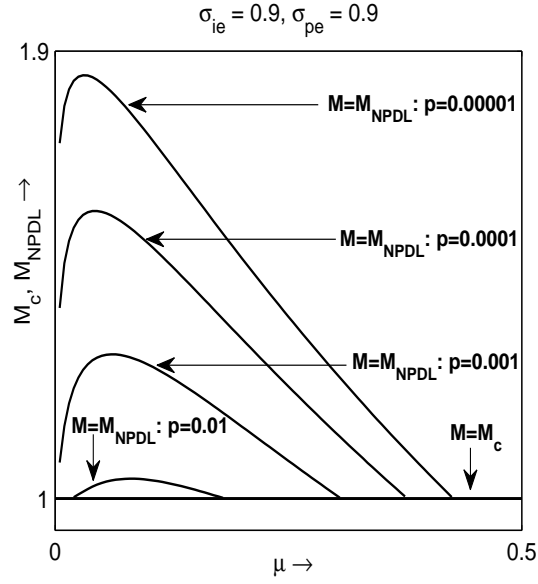}
  \caption{\label{NPDLs_for_different_p} Compositional parameter space with respect to $\mu$ showing the negative potential solitary structures for different values of $p$}.
\end{figure}

\begin{figure}
  \includegraphics{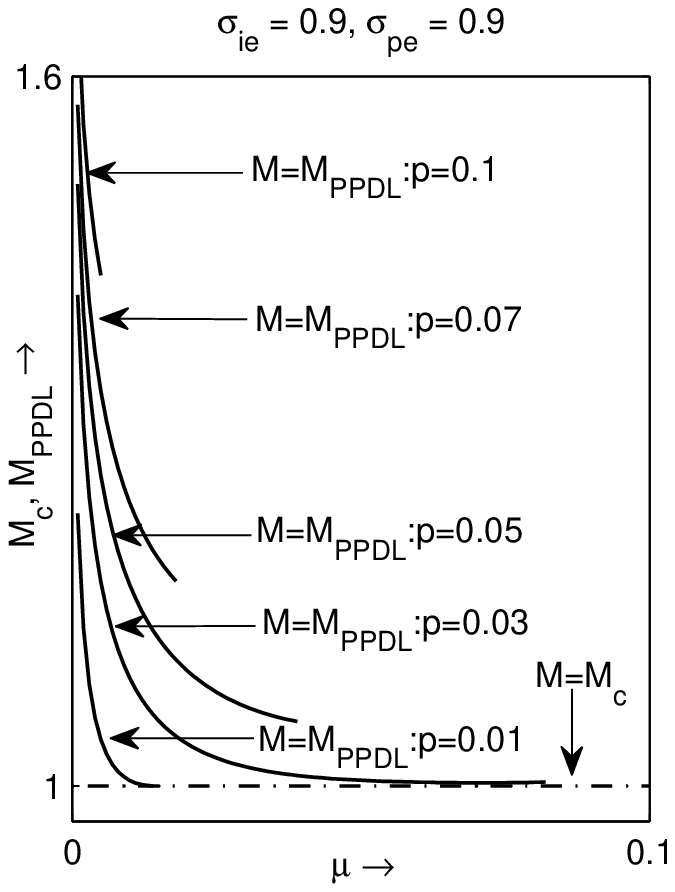}
  \caption{\label{PPDLs_for_different_p} Compositional parameter space with respect to $\mu$ showing the positive potential double layers and positive potential solitary waves restricted by positive potential double layers for different values of $p$}
\end{figure}

\begin{figure}
  \includegraphics{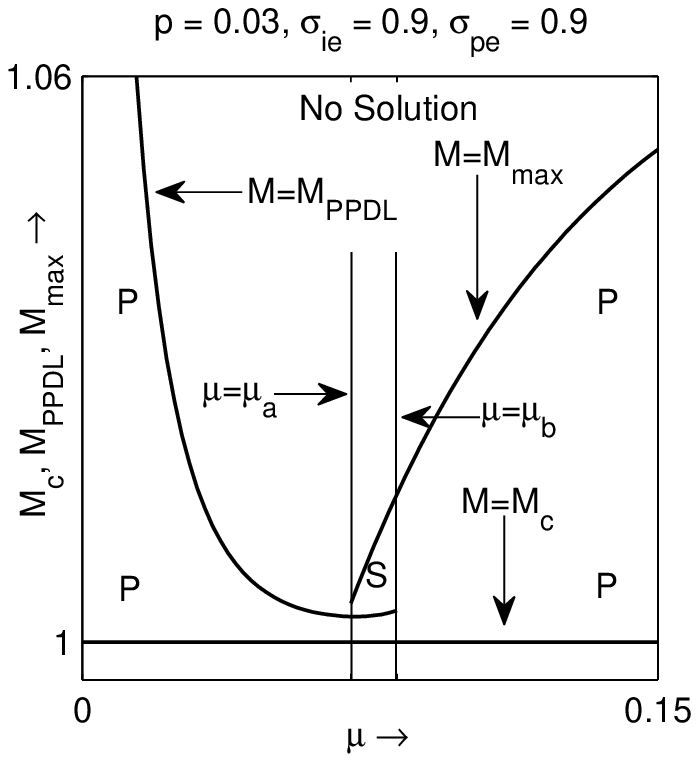}
  \caption{\label{sol_spc_wrt_mu_p=0_pt_03} Compositional parameter space with respect to $\mu$ for $p=0.03$, $\sigma_{ie}=0.9$ and $\sigma_{pe}=0.9$}
\end{figure}

\begin{figure}
  \includegraphics[width=8cm,height=8cm]{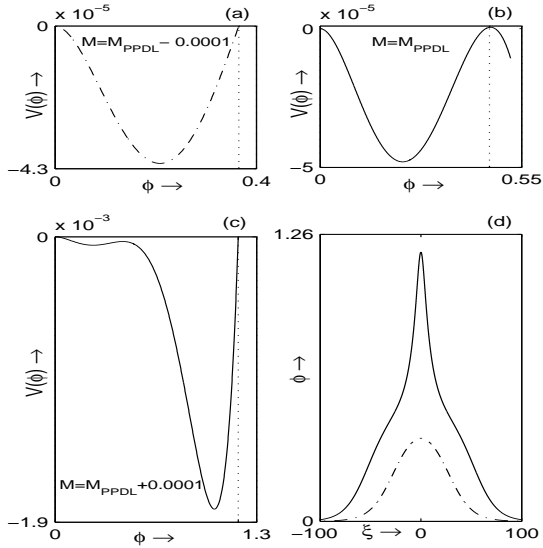}
  \caption{\label{profile_supersliton} $V(\phi)$ is plotted against $\phi$ for a value of $\mu$ lying in the interval $\mu_{a} < \mu < \mu_{b}$ for $p=0.03$: (a) $M=M_{PPDL}-0.0001$ (b) $M=M_{PPDL}$ (c) $M=M_{PPDL}+0.0001$. In (d), $\phi$ is plotted against $\xi$ for the same values of $\mu$ and $p$ for $M=M_{PPDL}-0.0001$ (dash-dot) and $M=M_{PPDL}+0.0001$ (solid).}
\end{figure}

\begin{figure}
  \includegraphics[width=8cm,height=8cm]{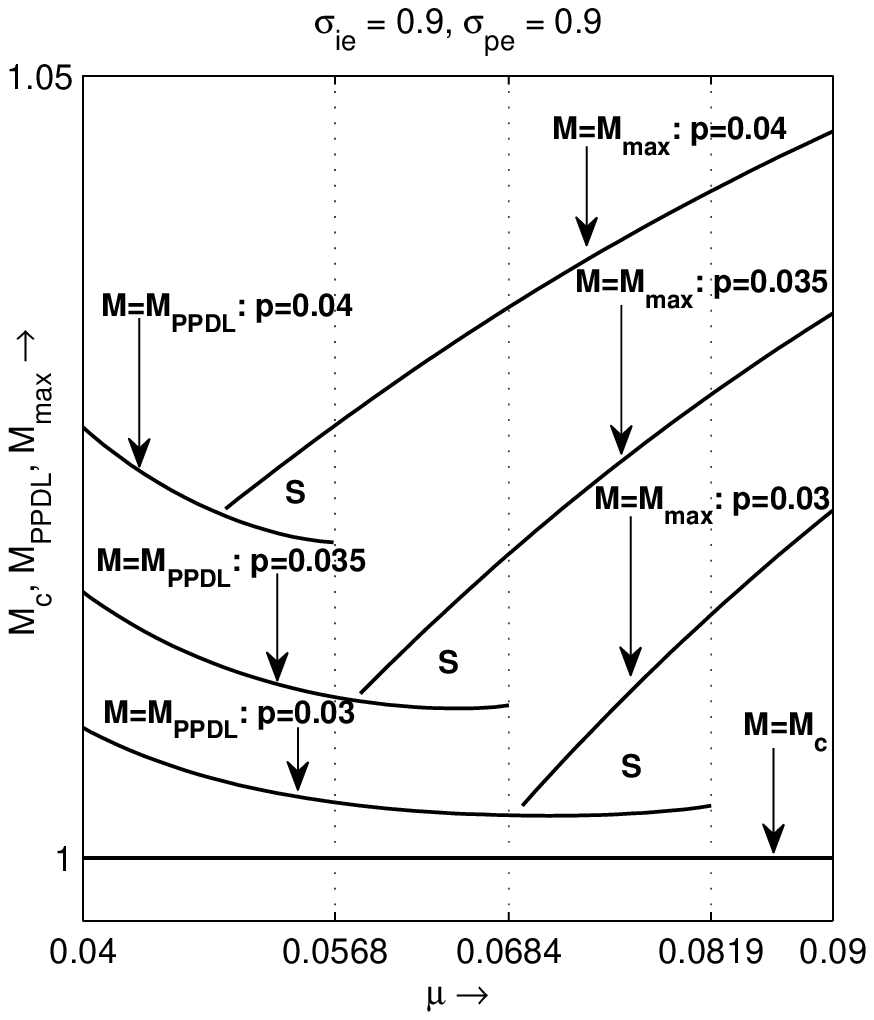}
  \caption{\label{supersolitons_different_p} Compositional parameter space with respect to $\mu$ showing the existence region of supersolitons for different values of $p$}
\end{figure}

\begin{figure}
  \includegraphics{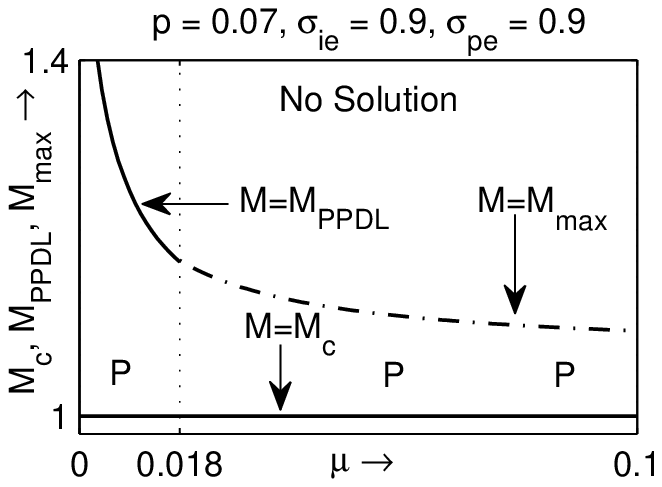}
  \caption{\label{sol_spc_wrt_mu_p=0_pt_07} Compositional parameter space with respect to $\mu$ for $p=0.07$, $\sigma_{ie}=0.9$ and $\sigma_{pe}=0.9$}
\end{figure}

\begin{figure}
  \includegraphics{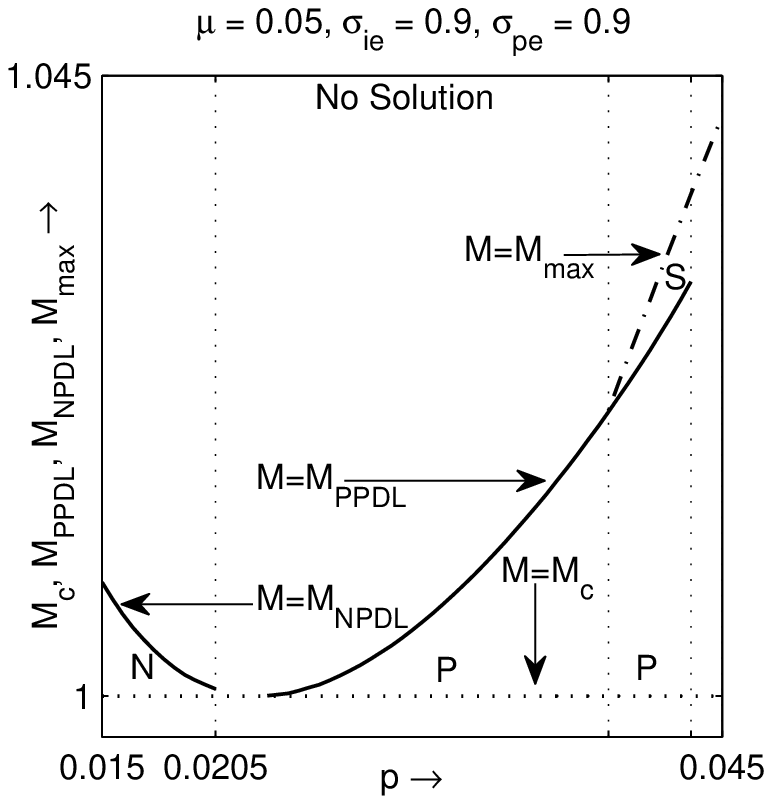}
  \caption{\label{sol_spc_wrt_p_mu=0_pt_05} Compositional parameter space with respect to $p$ for $\mu=0.05$, $\sigma_{ie}=0.9$ and $\sigma_{pe}=0.9$}
\end{figure}

\begin{figure}
  \includegraphics{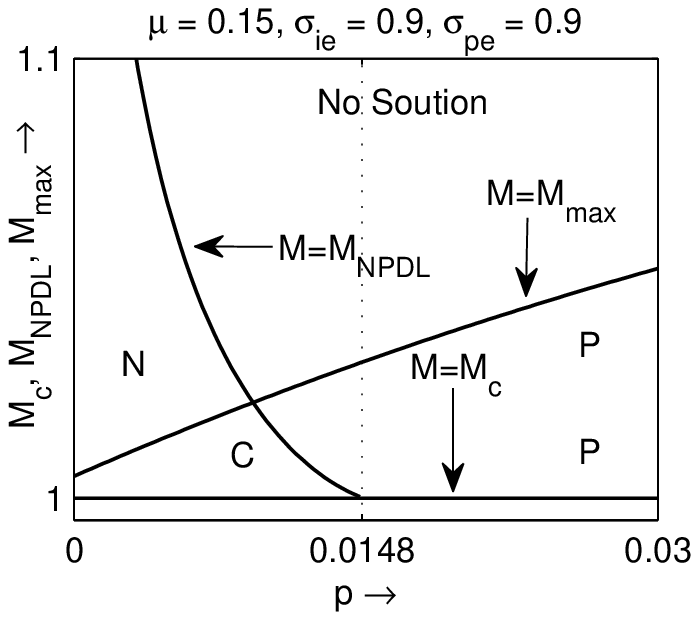}
  \caption{\label{sol_spc_wrt_p_mu=0_pt_15} Compositional parameter space with respect to $p$ for $\mu=0.15$, $\sigma_{ie}=0.9$ and $\sigma_{pe}=0.9$}
\end{figure}

In the present paper, a thorough investigation on the existence regions of the different dust ion acoustic solitary structures has been made in an unmagnetized dusty plasma consisting of negatively charged dust grains, adiabatic warm ions, and isothermally distributed electrons and positrons with the help of the compositional parameter space showing the nature of existence of different solitary structures, giving a special emphasis on the nature of existence of  supersolitons. If there exist solitons after the formation of double layer of same polarity for increasing values of the Mach number then the solitons are known as supersolitons and if the amplitude of any supersoliton is bounded then the supersolitons are known as bounded supersolitons. The Sagdeev potential technique has been used to study the arbitrary amplitude supersolitons. A general theory has been given on the existence of bounded supersolitons. If there exists a parameter regime for which the positive (negative) potential double layer is unable to restrict the occurrence of positive (negative) potential solitary waves then in this region of the parameter space, there exist positive (negative) potential solitary waves after the formation of positive (negative) potential double layer, and consequently, formation of positive (negative) potential supersoliton is confirmed. On the other hand, if the existence region or a portion of existence region of the total population of PPSWs (NPSWs) is separated by the curve $M=M_{PPDL}$ ($M=M_{NPDL}$)  or a portion of the curve $M=M_{PPDL}$ ($M=M_{NPDL}$), then for this particular portion of the curve $M=M_{PPDL}$ ($M=M_{NPDL}$), we have a positive (negative) potential solitary wave after the formation of positive (negative) potential double layer and as a result, we have a jump type discontinuity between the amplitudes of two types of solitary waves just before and just after the formation of double layer and consequently, formation of supersoliton is confirmed. In the present problem, we have found positive potential bounded supersolitons. The present dusty plasma system does not support any negative potential supersoliton. Not only $\mu$ and $p$ but also the parameter $\sigma_{pe}$ plays an important role for the formation of bounded positive potential supersolitons. Other than the formation of bounded positive potential supersolitons, we have found many interesting results regarding the existence of solitary structures from the qualitatively different compositional parameter spaces or solution spaces showing the nature of existence of different solitary structures with respect to any parameter of the system. One of the interesting observation is that the NPDLs occur in a deleted right neighbourhood of $p=0$ (see the solution spaces given in the figure \ref{sol_spc_wrt_p_mu=0_pt_05} and figure \ref{sol_spc_wrt_p_mu=0_pt_15}) whereas PPDLs exist in a deleted right neighbourhood of $\mu=0$ (see the solution spaces given in the figure \ref{sol_spc_wrt_mu_p=0_pt_01}, figure \ref{sol_spc_wrt_mu_p=0_pt_02}, figure \ref{sol_spc_wrt_mu_p=0_pt_03} and figure \ref{sol_spc_wrt_mu_p=0_pt_07}). These observations are very much consistent with the physical property for the formation of double layer which states that for the formation of NPDL (PPDL) solution at a point in the compositional parameter space the negative (positive) potential (absolute  value) must dominate the positive (negative) potential in a neighborhood of that point of the compositional parameter space and the  potential  difference must be maximum there. Another interesting observation is that for small value of positron concentration there is no qualitative change in solitary structures of DIA waves if very hot positrons are injected in the system. This fact is also consistent with the Energy Integral (\ref{energy_integral}).  Now, we want to summarize the main observations from the qualitatively different compositional parameter spaces with respect to $\mu$ for different values of $p$ and $\sigma_{pe}$ and also with respect to $p$ for different values of $\mu$.
\begin{enumerate}
  \item When there is no positron in the system, the system supports NPSW solution whenever $M>M_{c}$, i.e., the existence region of NPSWs is unbounded above whereas the existence region of PPSWs is bounded by the curves $M=M_{c}$ and $M=M_{max}$.
  \item Whenever a small amount of positron is injected in the system, the system supports NPDL solution along the curve $M=M_{NPDL}$ and consequently, the existence region of NPSWs has become bounded by the curves $M=M_{c}$ and $M=M_{NPDL}$. The existence region of NPSWs decreases for increasing values of positron concentration and finally, there exists a critical value of $p$ for which the system does not support any negative potential solitary structures (NPSWs and NPDLs). This fact can be efficiently described by the figure \ref{NPDLs_for_different_p}. The figure \ref{NPDLs_for_different_p} is the solution space with respect to $\mu$ for different values of $p$ showing the existence regions of NPSWs bounded by the curves $M=M_{c}$ and $M=M_{NPDL}$. If the negative potential solitary structures leave the system for a particular value of $p$ then it will never come back for further increment in positron concentration.
  \item When there is no positron in the system, the system does not support any PPDL solution. If we increase $p$ from $p=0$, then the system will support the PPDL solution whenever $p$ exceeds a cut off value $p_{1}$ (say). Then the existence region of PPSWs bounded by the curves $M=M_{c}$ and $M=M_{PPDL}$ will increase for increasing values of $p$ but here also there exists a cut off value $p_{2}$ (say) such that the existence region of PPSWs bounded by the curves $M=M_{c}$ and $M=M_{PPDL}$ increases for increasing $p$ for $p_{1}<p \leq p_{2}$ whereas the existence region of PPSWs bounded by the curves $M=M_{c}$ and $M=M_{PPDL}$ will decreases for increasing $p$ for $p_{2} \leq p < p_{3}$ and at $p=p_{3}$, the PPDL solution disappears from the system. This fact can be efficiently described by the figure \ref{PPDLs_for_different_p}. The figure \ref{PPDLs_for_different_p} is the solution spaces with respect to $\mu$ for different values of $p$ showing the existence regions of PPSWs bounded by the curves $M=M_{c}$ and $M=M_{PPDL}$. If the positive potential double layer leaves the system for a particular value of $p$ then it never comes back for further increment in positron concentration. But there always exist positive potential solitary waves bounded by the curves $M=M_{c}$ and $M=M_{max}$. So, although the PPDL solution disappears from the system for $p=p_{3}$, there always exist PPSWs bounded by the curves $M=M_{c}$ and $M=M_{max}$. In fact, for $p \geq p_{3}$, the system supports only PPSWs for any value of $\mu$.
  \item  From the previous discussions we have seen that if $p$ exceeds the cut off value $p_{1}$ but the value of $p$ is restricted by the inequality $p < p_{3}$ then we have two types of PPSWs: Type - I : PPSWs bounded by the curves $M=M_{c}$ and $M=M_{PPDL}$ which is same as First type PPSWs as defined earlier and Type - II : PPSWs bounded by the curves $M=M_{c}$ and $M=M_{max}$ which is not same as Second type PPSWs (see the solution space as given in figure \ref{sol_spc_wrt_mu_p=0_pt_01}, figure \ref{sol_spc_wrt_mu_p=0_pt_02}, figure \ref{sol_spc_wrt_mu_p=0_pt_03}, figure \ref{sol_spc_wrt_mu_p=0_pt_07}). Now if these two types of PPSWs are separated by a PPDL, i.e., if the existence region or a portion of existence region of the total population of PPSWs (i.e., union of Type - I and Type - II PPSWs) is separated by the curve $M=M_{PPDL}$  or the portion of the curve $M=M_{PPDL}$, then for this particular portion of the curve $M=M_{PPDL}$, we have a positive potential solitary wave after the formation of double layer (see the region marked as `S' of the solution space as given in figure \ref{sol_spc_wrt_mu_p=0_pt_03}). Now according to the property mentioned by \citet{das12}, we have a jump type discontinuity between the amplitudes of two types of solitary waves just before and just after the formation of double layer and consequently, formation of supersoliton is confirmed (see figure \ref{profile_supersliton}(d)). Now, the upper bound of PPSWs is either $M_{max}$ or $M_{PPDL}$. Obviously, in the existence region of positive potential supersolitons, we have $M_{PPDL}<M_{max}$ as positive potential supersoliton starts to exist just after PPDL. Therefore, for the bounded positive potential supersolitons, the following inequality holds good: $M_{PPDL}<M \leq M_{max}$, where $M$ is the Mach number corresponding to a positive potential supersoliton, i.e., a PPSW just after the formation of positive potential double layer. Figure \ref{supersolitons_different_p} shows that the existence region of the second type of PPSWs (i.e., positive potential supersolitons) decreases for increasing values of $p$ and finally disappears from the system if $p$ exceeds a certain cut off value.
  \item For fixed $\sigma_{pe}$, region of existence of PPSWs bounded by the curves $M=M_{c}$ and $M=M_{max}$ increases with increasing values of $p$.
  \item For fixed $\sigma_{pe}$, region of existence of PPSWs bounded by the curves $M=M_{c}$ and $M=M_{PPDL}$ increases for increasing $p$ for $p_{1}<p \leq p_{2}$ whereas it decreases for increasing $p$ for $p_{2} \leq p < p_{3}$.
  \item For fixed $\sigma_{pe}$, region of existence of NPSWs bounded by the curves $M=M_{c}$ and $M=M_{NPDL}$ decreases with increasing values of $p$.
  \item For fixed $\sigma_{pe}$, region of existence of supersolitons increases with decreasing values of $p$ up to a certain critical value of $p$ and then it decreases and finally it disappears from the system.
  \item For fixed $p$, region of existence of PPSWs bounded by the curves $M=M_{c}$ and $M=M_{max}$ decreases with increasing values of $\sigma_{pe}$.
  \item For fixed $p$, region of existence of PPSWs bounded by the curves $M=M_{c}$ and $M=M_{PPDL}$ decreases with increasing values of $\sigma_{pe}$.
  \item For fixed $p$, region of existence of NPSWs bounded by the curves $M=M_{c}$ and $M=M_{NPDL}$ increases with increasing values of $\sigma_{pe}$.
  \item For fixed $p$, region of existence of positive potential supersolitons increases with increasing values of $\sigma_{pe}$ up to a certain critical value of $\sigma_{pe}$ and then it decreases and finally it disappears from the system.
  \item From the solution space with respect to $\mu$, we have seen that for a particular $\mu$, it is not possible to get coexistence of NPSWs or NPDL and positive potential supersoliton whereas from the solution space with respect to $p$, we have seen that for a particular $p$, it is not possible to get coexistence of NPSWs or NPDL and positive potential supersoliton.
  \item From figure \ref{sol_spc_wrt_p_mu=0_pt_05}, it is clear that for a particular value of $p$, occurrence of NPDL does not imply the occurrence of positive potential supersoliton and conversely, occurrence of positive potential supersoliton does not imply the occurrence of NPDL, i.e., in any case, simultaneous occurrence of NPDL and positive potential supersolton is not possible. In fact, we have seen earlier that positive potential supersoliton comes into the picture after the disappearance of NPDLs for the increasing ( decreasing ) values of $p$ ($\sigma_{pe}$) whereas NPDL comes into the picture after the disappearance of positive potential supersolitons for the increasing (decreasing) values of $\sigma_{pe}$ ($p$).
  \item NPDL exists for small values of $\mu$ whereas PPDL exists for small values of $p$.
  \item For small value of $p$ and for very large value of $\sigma_{pe}$ ($\sigma_{pe}=100$, say), we have exactly same (qualitatively similar) type of solution space as given in figure \ref{sol_spc_wrt_mu_p=0}, i.e., the solution space for p=0. This fact is consistent with the Energy integral (\ref{energy_integral}).
\end{enumerate}

To conclude, we want to mention that the investigations made in the present paper through the qualitatively different compositional parameter spaces or solution spaces are helpful for the understanding of different nonlinear electrostatic structures of the present plasma system for the physical parameters of our interest. Our investigations are useful in understanding the properties of nonlinear dynamics of ion acoustic perturbations that may appear in astrophysical plasmas, such as those in the early universe and active galactic nuclei. In the present paper, we have not investigated the nonlinear behaviour of DIA waves when $U=c_{D} \Leftrightarrow W''(0)=0 \Leftrightarrow V''(0)=0 \Leftrightarrow M=M_{c}$, i.e., when the velocity of the wave frame is equal to the linearized velocity of the DIA wave for long wave length plane wave perturbation. The study of the dust ion acoustic solitary structures at the acoustic speed ($c_{D}$) in such plasmas is under consideration and the results shall be reported elsewhere.

\acknowledgments One of the authors (Ashesh Paul) is thankful to the Department of Science and Technology, Govt. of India, INSPIRE Fellowship Scheme for financial support.

\bibliography{aps_apadab}% Produces the bibliography via BibTeX.

\newpage

\begin{figure}
  \includegraphics{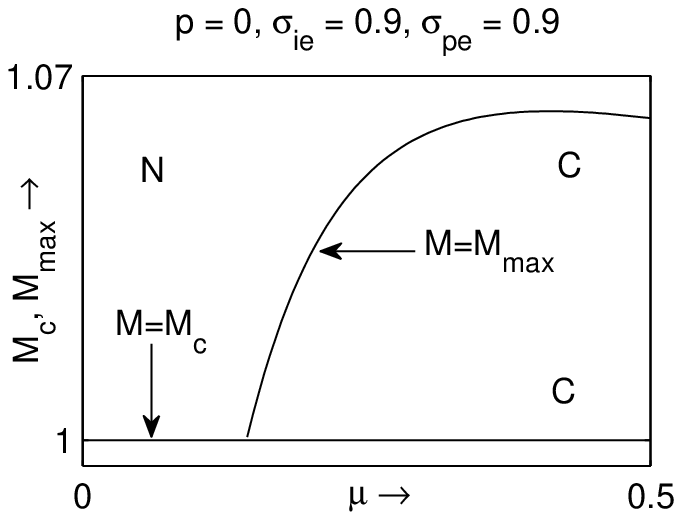}
  \caption{\label{sol_spc_wrt_mu_p=0} Compositional parameter space with respect to $\mu$ for $p=0$, $\sigma_{ie}=0.9$ and $\sigma_{pe}=0.9$}.
\end{figure}

\end{document}